\documentclass[twocolumn]{svjour3}      
\smartqed  
\usepackage{graphicx}
\usepackage{mathptmx}      
\usepackage{mathrsfs}      
\usepackage{amsmath}		
\usepackage{amssymb}
\usepackage{multirow}
\usepackage{booktabs}
\usepackage{lineno}
\usepackage{bm}
\usepackage{pifont}
\journalname{Computational Particle Mechanics}
\usepackage{hyperref}
\hypersetup{
    colorlinks=true,
    linkcolor=blue,
    filecolor=blue,      
    urlcolor=blue,
    citecolor=blue,
    bookmarks=true
}
\newcommand*{\orcid}{\protect\includegraphics[scale=0.5]{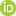}} 
\usepackage{tikz}

\begin{document}
\title{High compression of granular assemblies of brittle hollow tubular particles}
\subtitle{Investigation through a 3D discrete element model}
\author{
M.~Stasiak \textsuperscript{a}\href{https://orcid.org/0000-0001-7623-2308}{\orcid} \and
G.~Combe \textsuperscript{a}\href{https://orcid.org/0000-0002-8633-0793}{\orcid} \and
V.~Richefeu \textsuperscript{a}\href{https://orcid.org/0000-0002-8897-5499}{\orcid} \and
G.~Armand \textsuperscript{b} \and
J.~Zghondi \textsuperscript{b}
}
\authorrunning{Stasiak\href{https://orcid.org/0000-0001-7623-2308}{\orcid} \textit{et al.}} %
\institute{
\textsuperscript{a} Institute of Engineering and Management, Univ. Grenoble Alpes, 3SR, 38000 Grenoble, France.\\
\email{gael.combe@grenoble-inp.fr}  \\     
\textsuperscript{b} Andra, R\&D Division, Meuse/Haute-Marne Underground Research Laboratory, 55290 Bure, France.
}
\date{Received: date / Accepted: date} 
\maketitle
\begin{abstract}
This paper is devoted to the micro-mechanical origins of the high compressibility of brittle tubular particle assemblies. 
The material is extremely porous due to the presence of a large hole within the tube-shaped particle.
The release of the inner void, protected by a fragile shell, gives the material a very strong ability to compress. The compressive response is investigated by means of the Discrete Element Method, DEM, using crushable-elements.
To address the complexity of the model, a step-by-step break-down is developed.

The paper comprises the comparison of the numerical results with both results obtained by the authors and existing experiments.
With the insights provided by the DEM, we have sought to better understand the phenomena that originate at the grain scale, and that govern macroscopic behaviour.
Grain breakage was proven to control the compressive behaviour, and thus, the importance of internal pores dominates the inter-particle voids. 
Then, a novel concept of compressibility analysis has been proposed using the separation of the double porosity and the quantification of the pore collapse through primary grain breakage. Finally, a general, geometrical development of a semi-analytical model has been proposed aiming the prediction of the evolution of double porosity \textit{vs} axial strain.
\end{abstract}

\section{Introduction\label{secIntro}}

\begin{figure}[htb]
\centering
\includegraphics[width=0.98\linewidth,trim={0 20 0 0},clip]{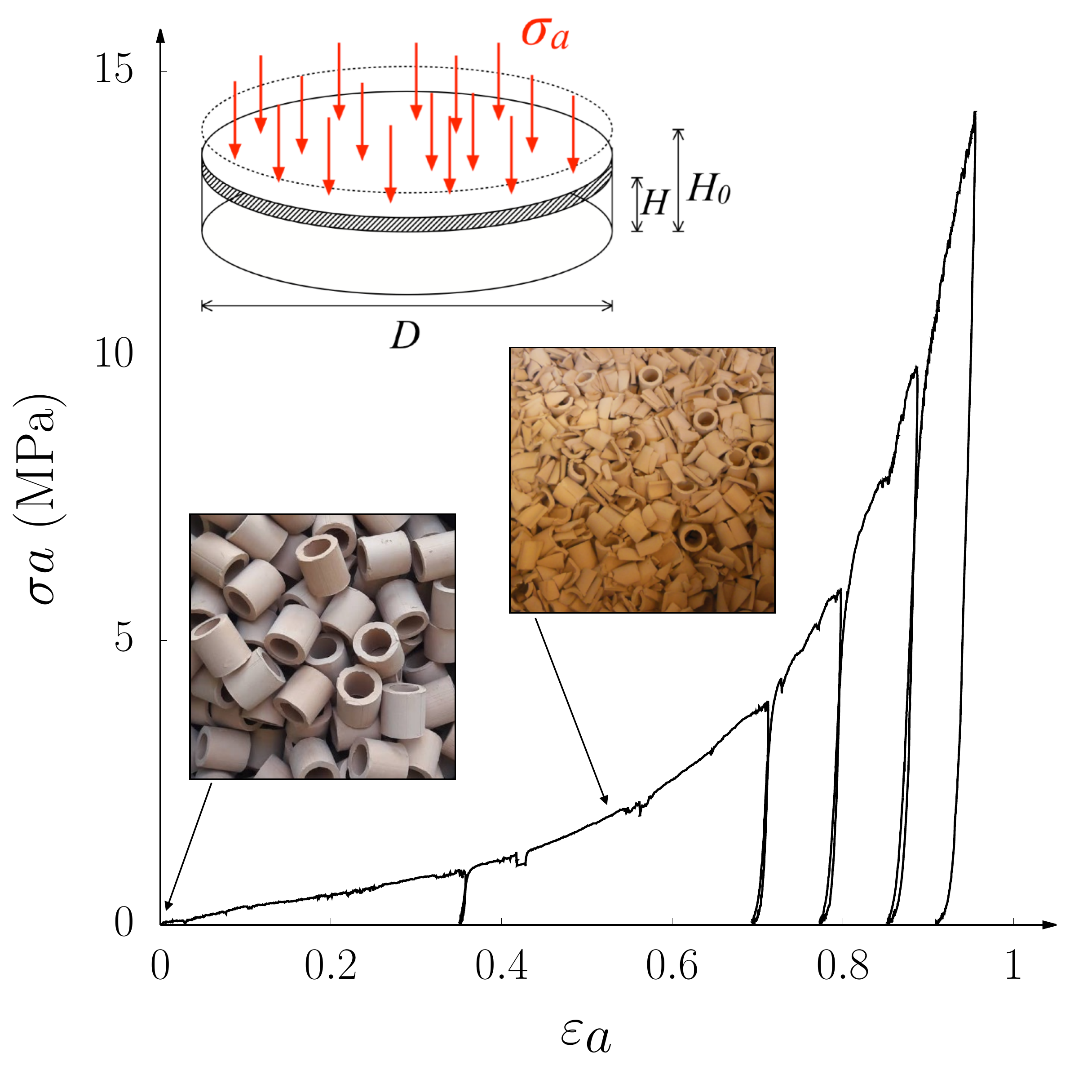}
\caption{Typical compressive response of a \textit{shell} assembly (Figure~\ref{FigCluster}a) to oedometeric compression: $\sigma_a$ -- uniaxial, imposed load and $\varepsilon_a$ -- compressible deformations.}
\label{FigExpMechBeh}
\end{figure}

\begin{figure}[htb]
\centering
\includegraphics[width=0.45\linewidth,trim={0cm 0cm 350 80},clip]{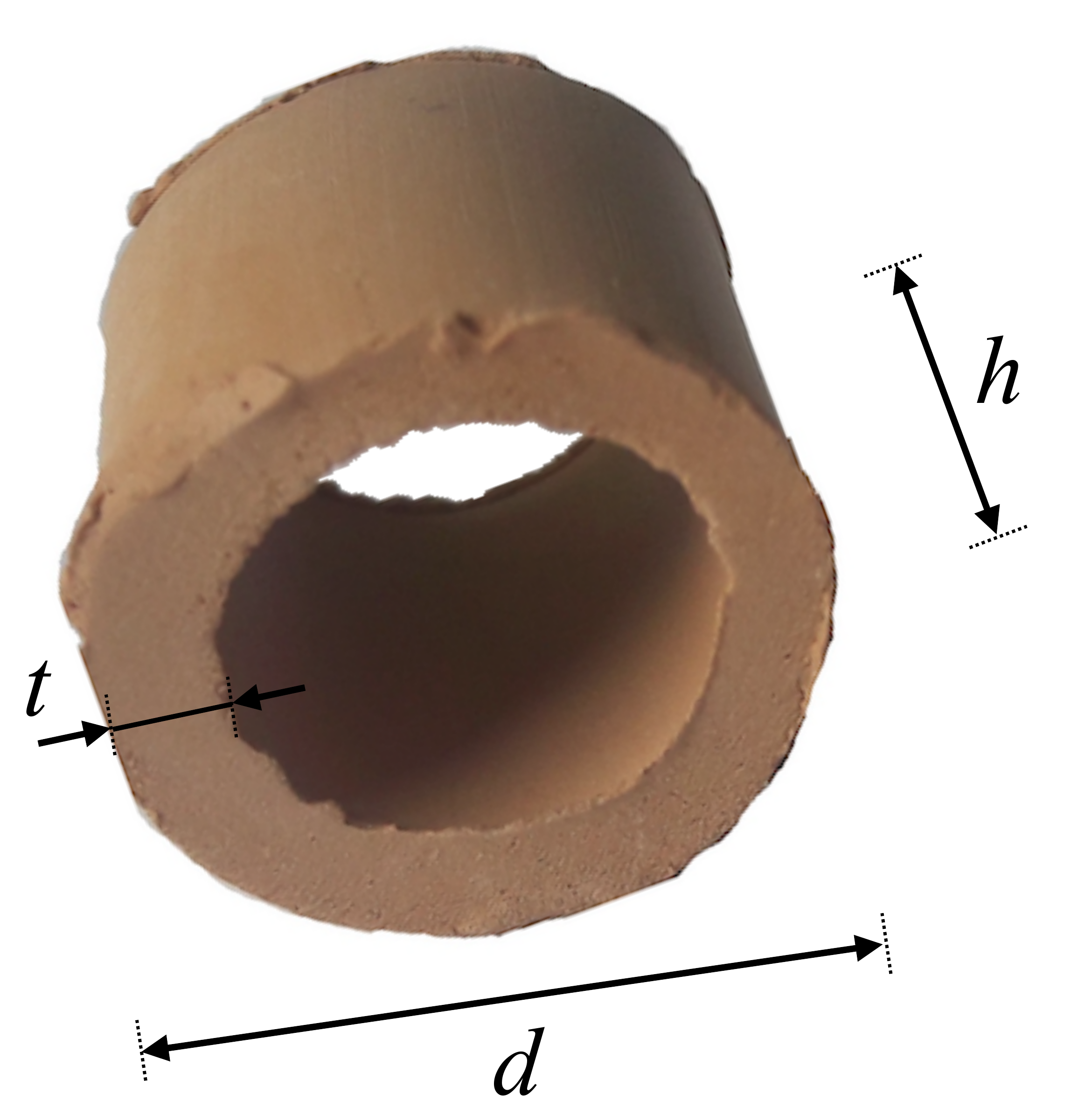}
\includegraphics[width=0.45\linewidth,trim={0cm 0cm 325 80},clip]{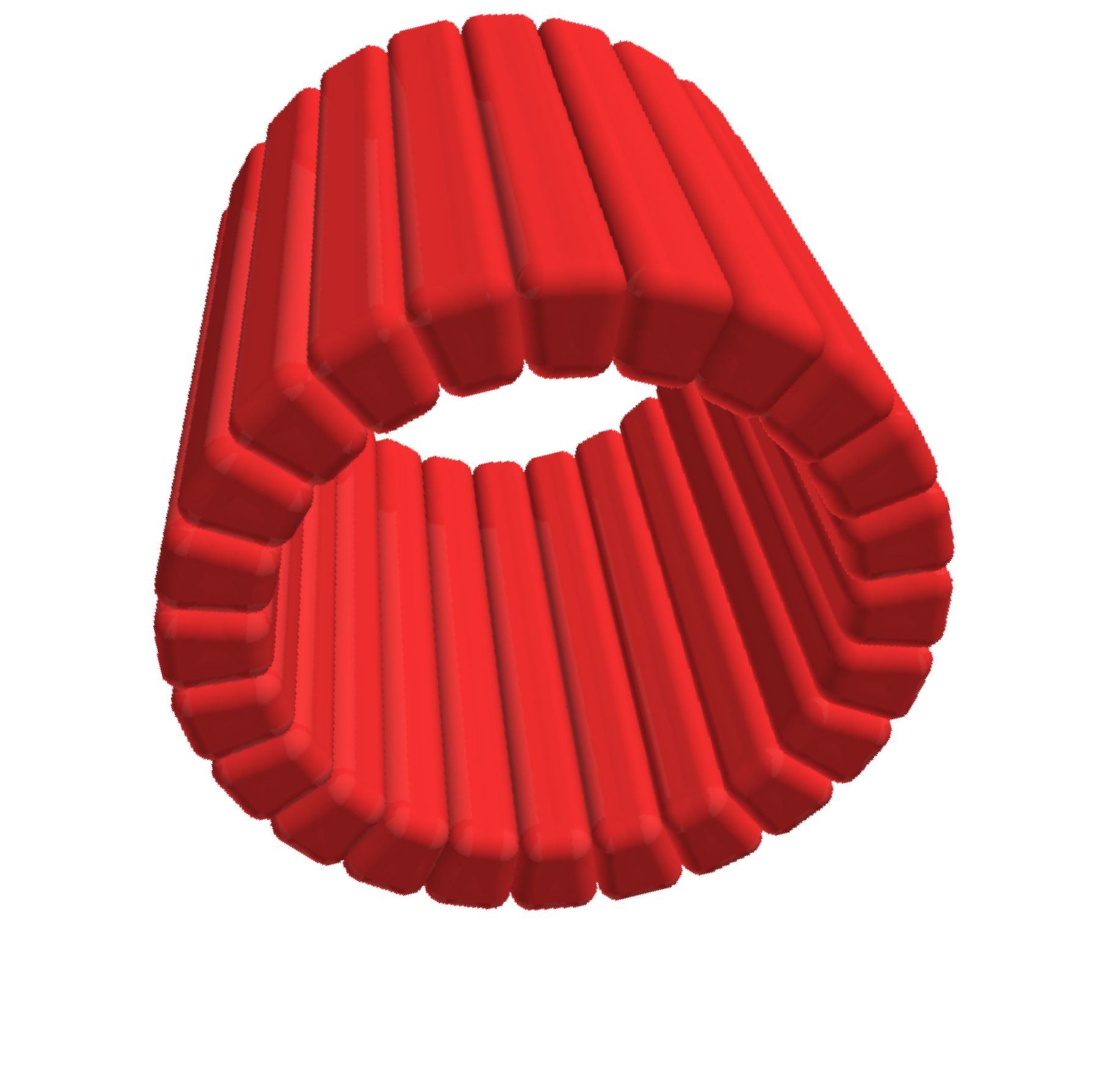}
\begin{flushleft} \hspace{1.4cm} {\large(a)} \hspace{4.cm} {\large(b)} \end{flushleft}
\caption{A brittle, crushable particle with highly porous geometry: (a) a tube-shaped \textit{shell} is described by diameter $d$, height $h$ and thickness $t$ such that $d\simeq  h\simeq 20$~mm and $t\simeq 2.4$~mm, (b) {a \emph{cluster},  a numerical clone of the shell, with same dimensions} and made of 24 rigid \textit{sectors} linked together as shown in Figure~\ref{FigSector}.}
\label{FigCluster}
\end{figure}

The research presented in the current paper originates from an engineering concept of the mono-block compressible arch-segment (US Patent pending on ``Voussoir Monobloc Compressible'', VMC, developed jointly by CMC and Andra \cite{Patent}).
VMC is an innovative technique of fabricating the tunnel lining. This pre-casting technology combines a conventional reinforced concrete segment with a compressible layer on its outer side (extrados) \cite{Andra2016,Andra2017,Zghondi2018,Taherzadeh2019}. 
In the engineering practice, there exist some cases in which the host rock around tunnel was partially replaced with a compressible material that prevents an overloading of already existing tunnels \cite{Fernandez2019}. Such approach treats the tunnel and the compressible material as separate systems, whereas in Andra's double-layer lining the two systems cooperate closely.
Another innovation of the VMC design concerns the compressible layer itself. Instead of a highly compressible material, like foam, a cemented assembly of highly porous and quasi-fragile particles was envisioned.  
From a mechanical point of view, the influence of the weak cement bonds between the particles is essentially dominated by the characteristics of the granular core. The key benefit of introducing the granular material is related to the charge transfer that it enables \cite{Chevalier2012}: 
a local load can be minimised by spreading it on the extrados surface of the reinforced concrete segment.
In addition, under sufficient load, the layer will experience some compressible deformations allowing for a stress dissipation (Figure~\ref{FigExpMechBeh}).
As long as the outer layer exhibits its compressible capacities, it limits the load transferred on the inner layer.
The quasi-fragile particles were manufactured from the excavated COx clay stone, first dried then baked at high temperature, and therefore have brittle properties.
The compressible response (Figure~\ref{FigExpMechBeh}) is predominantly achieved by closing the pores. The large internal porosity contributes significantly to a highly compressible mechanical response, \textit{e.g.}, \cite{Guida2018}.
This material owes its high porosity to the tubular geometry of the constituent particle, called \emph{shell} (Figure~\ref{FigCluster}a).
The geometry was specially selected for this application, such that the shell-scale porosity, \textit{i.e.}, volume of internal void to total volume of particle, reaches 50\%.

First purpose of this work was to design a numerical model capable of simulating shell breakage.
{For this, methods based on peridynamics, \textit{e.g.} \cite{zhu2019}, or the material points method, \textit{e.g.} \cite{Xiao2021}, would be suitable but expensive in computation time. Another approach like Discrete Element Method, DEM, as it was introduced in the 80s \cite{Cundall1979}, is also particularly well suited for breackage granular materials.}
A couple of DEM techniques can include the capability of the particle to break:
fragment substitution, FS, \textit{e.g.}, \cite{Tsoungui1999,Ben-Nun2010,Bono2016}, and Bonded Particle, BP,  \textit{e.g.}, \cite{Thornton1996,Bolton2008,Wang2011}. 
In FS methods, the breakage criterion is defined as a function of the critical load that particle cannot withstand. Once broken, the original particle is replaced by a set of new particles. The replacement mode is usually structured under the condition of a self-similarity. In other words, the replacement particles keep the same shape but are smaller in size. Self-similarity suits a material like sand but is hardly applicable in the case of tubular shells. It is then more convenient to use BP approach that connects sub-particles by means of the cohesive links. The breakage is verified between sub-particles, thus it requires knowledge of a material strength. 

A rigid clump of elements can in detail reproduce highly complex shapes \cite{Ferellec2008}. 
In the field of geomechanics, this technique was found considerably useful in the investigations of the particle shape, for example aiming the concavity effect \cite{Szarf2011,Azema2013}. 
This work also took advantage of clumping method when constructing a non-spherical polygonal-like \emph{sector}. To generate the tube-shaped geometry (Figure~\ref{FigCluster}b), the sectors were linked together using BP method. Section \ref{secModel} elaborates on how the BP Model was adapted to the case of the cylindrical shells. The paper is then continued with shell-scale investigations, both numerical and experimental, in Section~\ref{secPramMicro}. This part is dedicated to the identification of discrete parameters and a calibration of the breakage model. Section~\ref{secExpeCharacterisation} is a tribute to the true packing with the focus being redirected to the internal state variables. It mostly discusses shell orientation: the theoretical approach and the results of an experimental characterisation with X-ray acquisitions. Sensitivity analysis presented in Section~\ref{secParamStudy} gives a better understanding of how the discrete parameters as well as the initial state affect the macroscopic behaviour to oedometric compression. At this point, the complexity of the model, the necessary compromises and the validity point will be also discussed. This section also begins the discussion over high compressibility exposed by the stress-strain curve (with a reference to the experiment). Yet, the analysis of compressibility is fully completed in Section~\ref{secDiscus}. To finish, a geometrical and analytical development of a prediction model incorporating the breakage phenomenon is proposed. Finally, a brief summary is given at a very end, in Section~\ref{secEnd}.

\section{Model of crushable shell\label{secModel}}

{The experimental study of the behavior under oedometric compression of tube shaped clay-shell assemblies shows that before being completely crushed and reduced to powder ($\varepsilon_a =  0.9$ in Figure~\ref{FigExpMechBeh}), the particles break essentially radially, up to  $\varepsilon_a \simeq  0.6$, insets in Figure \ref{FigExpMechBeh}.}

\begin{figure}[htb]
\centering
\includegraphics[width=0.7\linewidth,trim={0 100 400 60},clip]{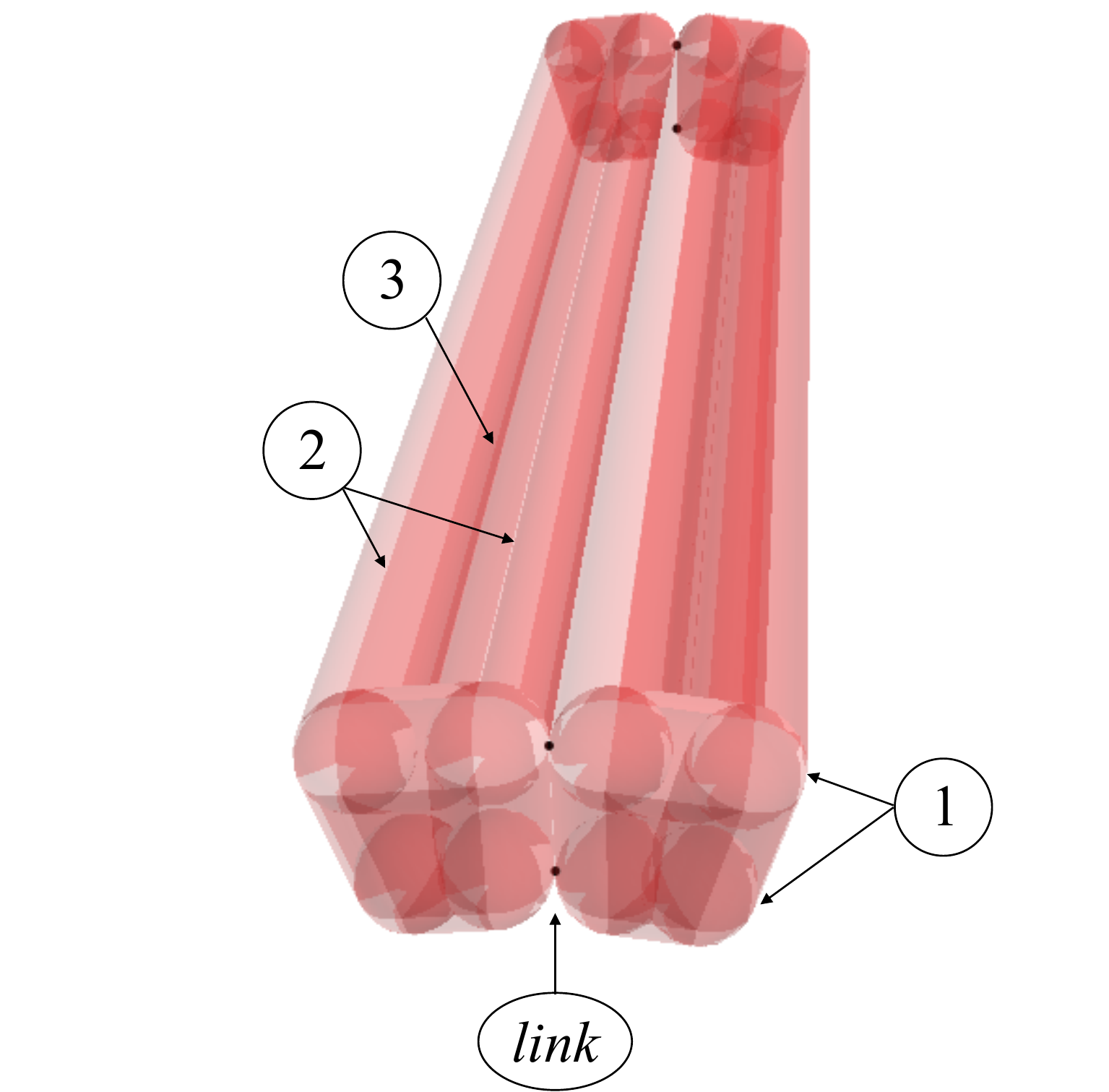}
\caption{Two cohesively linked sectors. A \textit{sector} is a rigid clump of the elementary solid figures (spheres \textcircled{1}, cylinders \textcircled{2} and rectangular prism \textcircled{3}) that spatially form a \textit{sphero-polyhedron}. The black dots indicate the \emph{links} that act as an elastic glue while they are not broken (Figure~\ref{FigForceLawGlue}). Four sphere-to-sphere links are accounted for each sector interface. {The segment size is $h=20~mm$ and $t=2.4~mm$, see Figure~\ref{FigCluster}. The segment width depends on the number $N^{\star}_{circ}$ of segments used to model the shell}. }
\label{FigSector}
\end{figure}

To simulate the breakage of the \emph{shell} (Figure~\ref{FigCluster}a), this model was based upon the Bonded Particle (BP) approach -- a DEM technique \emph{clustering} smaller, rigid sub-particles into a ``parent'' particle.
Therefore, in the current paper a numerical, tube shaped, breakable particle is referred to as the \emph{cluster}, while term \emph{sector} stands for its constituent (Figure~\ref{FigCluster}b).
As presented in Figure~\ref{FigSector}, each sector has been generated as a polyhedron-like shape by clumping 3D solid figures (sphere, cylinder and rectangular prism) rigidly\footnote{lack of relative movement between the solid figures} together,and thus, preventing it from any deformations.
Sectors of the same parent cluster are joined by means of the distinctive connections, hereinafter called \emph{links}.
For the sake of simplicity, the link is only located in-between spheres that are positioned at the sector ``corners'' (Figure~\ref{FigSector}). 
Relative movements evolve in all four links aligned on an interface between two sectors, and thus, those links develop the elastic forces as follows:
\begin{equation} \label{eqLinkForces}
\begin{pmatrix} 
f_{I} \\ 
f_{II}
\end{pmatrix} = -
\begin{pmatrix} 
k_{I} & 0 \\
0 & k_{II} 
\end{pmatrix}
\cdot
\begin{pmatrix} 
\delta_{I} \\
\delta_{II} 
\end{pmatrix}
\end{equation}
where $k_{\bullet}$ are the link stiffnesses, and $\delta_{\bullet}$ are the relative displacements between two spheres at there touching point. In other words, $\delta_{\bullet}$ can be considered as shortening or elongation of the local links in corresponding direction. The subscripts \textit{I} and \textit{II} refer to the fracture mode, but can also be seen as the normal and tangential direction of the link.

\begin{figure*}[htb]
\begin{flushleft} \hspace{2cm} (a) \hspace{4.5cm} (b) \hspace{4.5cm} (c)  \end{flushleft}
\centering
\includegraphics[width=0.3\linewidth,trim={0 4cm 15cm 0},clip]{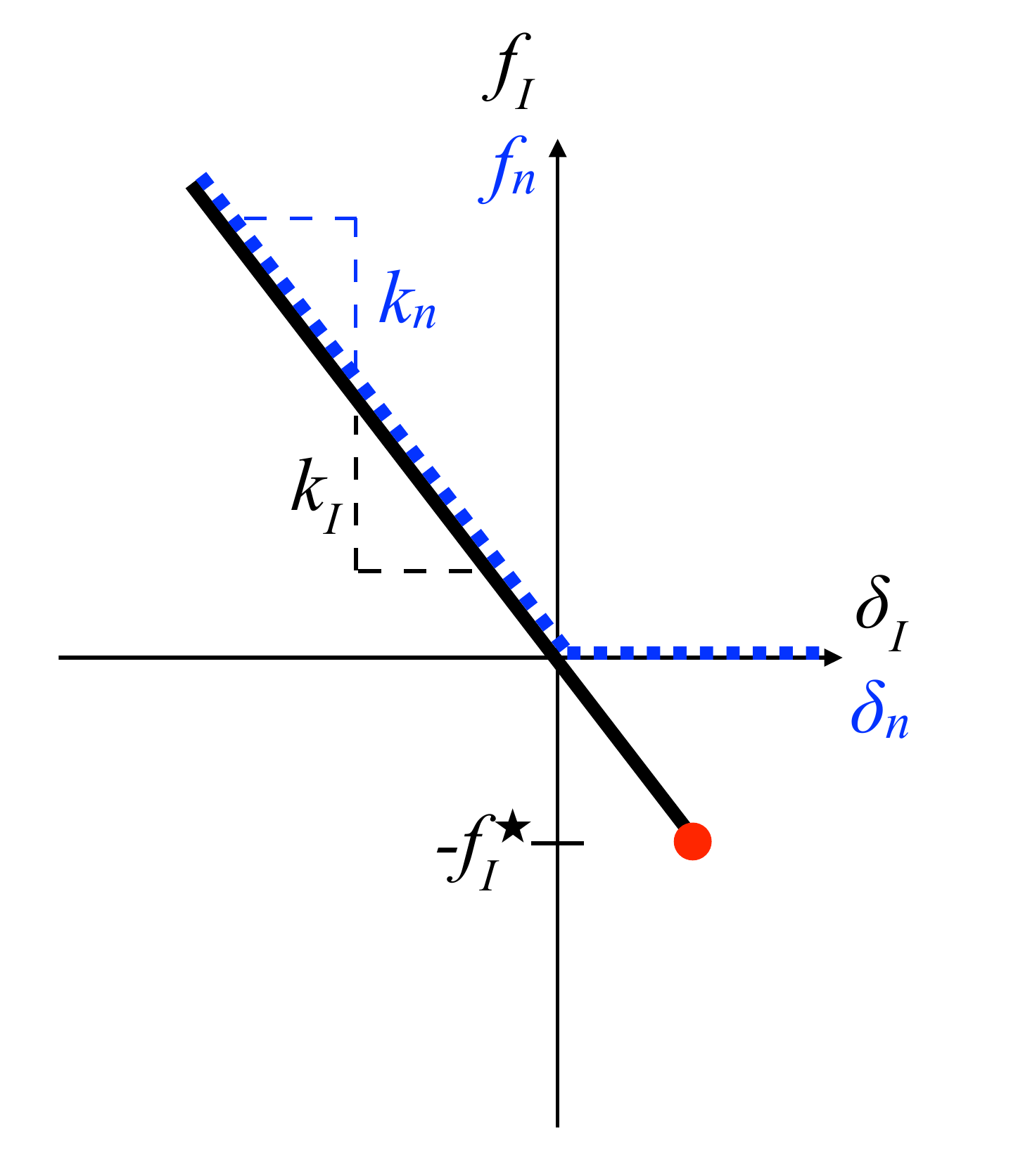}  
\includegraphics[width=0.3\linewidth,trim={0 4cm 15cm 0},clip]{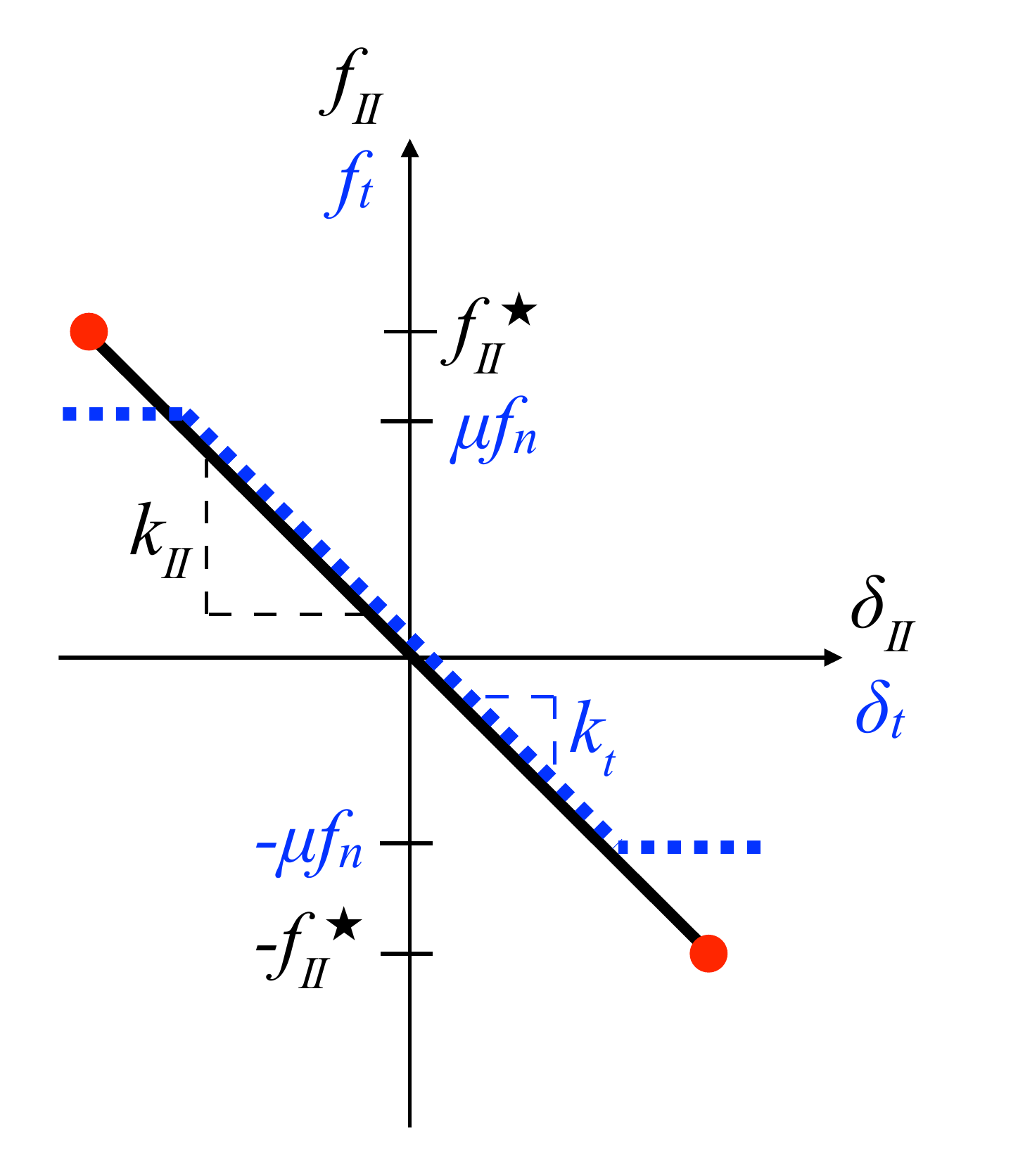} 
\includegraphics[width=0.3\linewidth,trim={0 4cm 15cm 0},clip]{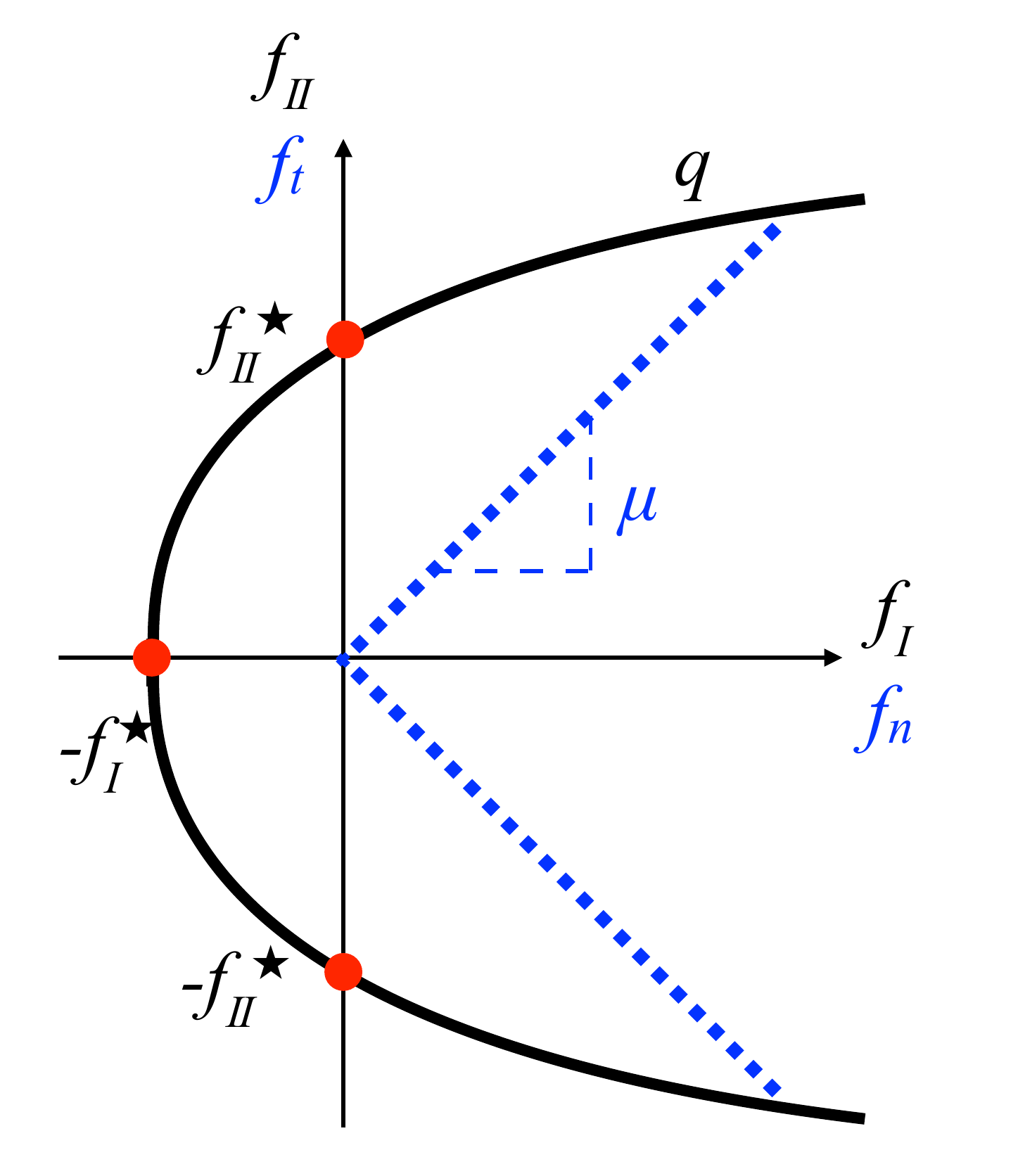} \\
\includegraphics[width=0.35\linewidth,trim={0 15cm 8.5cm 0},clip]{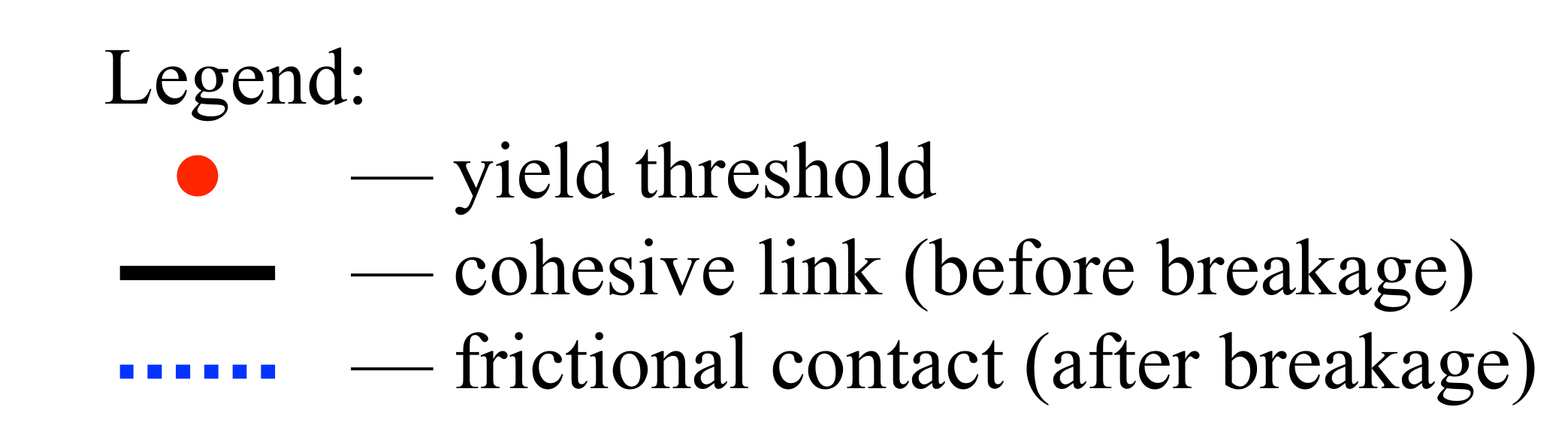} \hspace{10.5cm}
\caption{
Graphic representation of the force laws for the links (black lines) and for the frictional contacts (dashed blue lines): (a) loading in mode-\textit{I}  (normal direction); (b) loading in mode-\textit{II} (tangential direction). (c) A 2D representation of the critical state for links in $f_I - f_{II}$ space (black curve), and for frictional contacts in $f_t - f_n$ space (dashed blue lines). The latter envelopes are (\textit{i}) the yield function of Equation~\ref{eqYieldFunction} for a given $q$, and (\textit{ii}) the Coulomb cone for a given friction $\mu$.}
\label{FigForceLawGlue}
\end{figure*}

To avoid pure mode-\textit{I} fracture ($f_{II}=0$~N), $f_{I}$ must not exceed a tensile yield threshold $-f_{I}^\star$ (Figure~\ref{FigForceLawGlue}a). The material also withstands shear of the pure mode-\textit{II} type ($f_{I}=0$~N), as long as the force $f_{II}$ remains in the $\pm f_{II}^\star$ range (Figure~\ref{FigForceLawGlue}b).
The yielding surface for any mixed loading (combined mode-\textit{I} and mode-\textit{II}) in the $f_{I}$--$f_{II}$ space, is defined by the following function:
\begin{equation} 
\label{eqYieldFunction}
\varphi = \dfrac{f_I}{-f_I^\star} + \left(\dfrac{\vert f_{II} \vert}{f_{II}^\star}\right)^{q} - 1
\end{equation}
where $q$ enables to adjust the surface between a bi-linear shape ($q=0$) and a semi-infinite rectangle shape when $q$ tends towards infinity. Figure~\ref{FigForceLawGlue}c shows an example of yield surface for $q$ in the order of 2.
All four links are broken as soon as in one of them the breakage criterion is met:
\begin{equation} 
\label{eqFailCriterion}
\varphi \geq0
\end{equation}
Once the link is yielded, it is irreversibly broken.
At this stage, frictional contact interactions become possible between the two sectors that have just been detached.

In case of frictional contact, the force-displacement relations used are quite classical, and can read as
\begin{equation}
    f_n = 
    \begin{cases}
    -k_n \delta_n & \text{if}\ \delta_n < 0 \\
    0 & \text{otherwise}
    \end{cases}
\end{equation}
and
\begin{equation}
    \Delta f_t = 
    \begin{cases}
    -k_t \dot{\delta}_t \Delta t & \text{if}\ f_t < \mu f_n \\
    0 & \text{otherwise}
    \end{cases}
\end{equation}
where $f_n$ is a normal repulsive force proportional to the normal distance $\delta_n$ in case of overlap, and the increment of tangential force $\Delta f_t$ is proportional to the increment of sliding distance (that is the sliding velocity $\dot{\delta_t}$ multiplied by the time step $\Delta t$) as long as $f_t$ remains below $\mu f_n$.
As a result, a contact involves only compressive $f_n$ as demonstrated in Figure~\ref{FigForceLawGlue}a. The tangential force is constrained to remain within the range $\pm \mu f_n$; Figure~\ref{FigForceLawGlue}b. This leads to a bi-linear yield function (Coulomb cone) ruled by a given friction coefficient $\mu$; Figure~\ref{FigForceLawGlue}c.

In discrete element simulations of quasistatic regimes \cite{combe2003}, energy dissipation is a major concern \cite{Atman2009}. In our situation, the fracture of the brittle shells is characterised by high energy releases, so the efficiency of energy dissipation becomes even more critical. To handle it, two means are employed. The first is a viscous force added to the contact normal force, such that the damping is at 95\% of its critical state corresponding to perfectly inelastic collisions. The second means of damping is not strictly physical but rather numerical. It is the non-local damping which has been set to 30\% as prescribed by their designers \cite{Cundall1987}. The purpose of its use is to minimize the surges of kinetic energy induced by sudden link failures.
These two damping parameters were kept identical for all simulations, and we obviously ascertained that they did not significantly affect the macroscopic mechanical responses.

The numerical part of the work has been conducted using a parallelised tool named \texttt{Rockable}\footnote{In house numerical code developed by V. Richefeu \href{https://orcid.org/0000-0002-8897-5499}{\orcid}, in collaboration with the authors of this paper} which algorithm operates on complex, sphero-polyhedral shape  features (here they are sectors). Regarding the force laws, one needs to specify 8 parameters:
\begin{itemize}
\item[\ding{217}] 5 for the links (ruling the breakage): $k_{I}$, $k_{II}$, $f_{I}^\star$, $f_{II}^\star$ and $q$,
\item[\ding{217}] 3 related to the frictional contacts (between detached sectors or clusters): $k_{n}$, $k_{t}$ and $\mu$.
\end{itemize}
In the following Section~\ref{secPramMicro}, their determination and/or tuning will be discussed.
Notice that the adjustment of some parameters ($k_{n}$, $k_{t}$ and $\mu$) require investigations on assemblies  
so that clusters can interact. This aspect is to be discussed further in the section~\ref{secParamStudy}.
 
\section{Assessment of shell mechanical properties\label{secPramMicro}}

The model requires a calibration. This process of parameter tuning, which is essential to the reliability of the model, is not a straightforward task of bridging quantities at the micro and macro scales.
We are focusing here on the breakage of a single shell. The strength of the constituent material of the shell must be considered in the evaluation of the threshold forces of Equation~\eqref{eqYieldFunction}.
The present modelling is such that the parameters $f_{I}^\star$ and $f_{II}^\star$ monitor the resistance of the shell in combination with the chosen sectioning (\textit{i.e.}, the way the cluster is partitioned). In other word, the threshold forces are structural parameters and they are in no way constitutive parameters. Therefore, experiences of breaking a single shell have been relied on.

\begin{figure}[htb]
\centering
\includegraphics[width=0.98\linewidth,trim={0 0 200 0},clip]{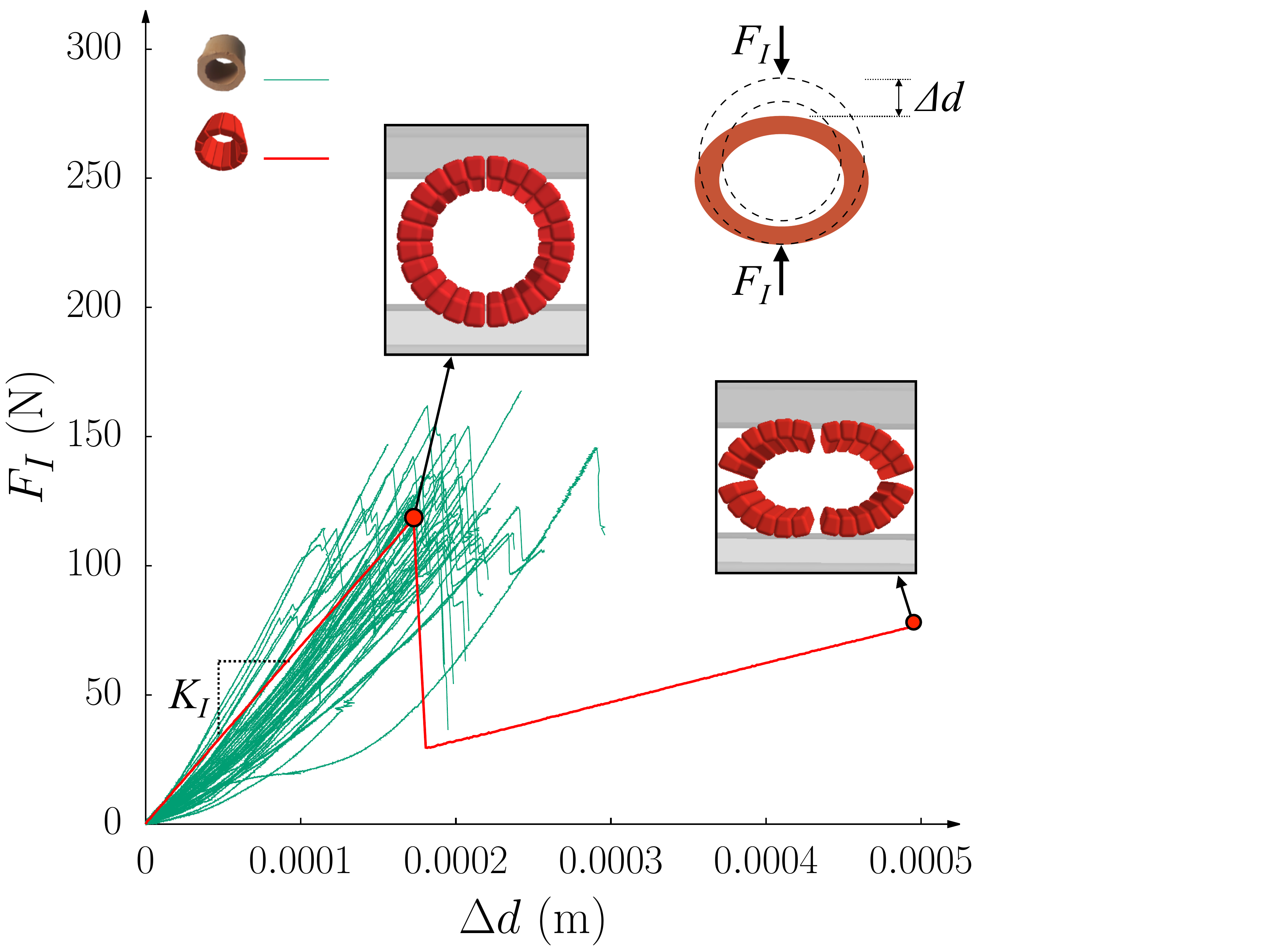}
\caption{Uniaxial Radial Compression (top right scheme). A variability of the experimental behaviour is shown by the green curves (51 tests). The red line is the numerical reflection of the  mechanical behaviour ruled by shell stiffness $K_I$ up to the primary breakage. The localisation of primary  (vertical) and secondary (horizontal) cracks are presented in the two insets.}
\label{FigExpNumBT}
\end{figure}

\subsection{Mode-\textit{I} failure ($k_{I}$, $f_{I}^\star$)}

\begin{figure}[htb]
\centering
\includegraphics[width=0.98\linewidth,]{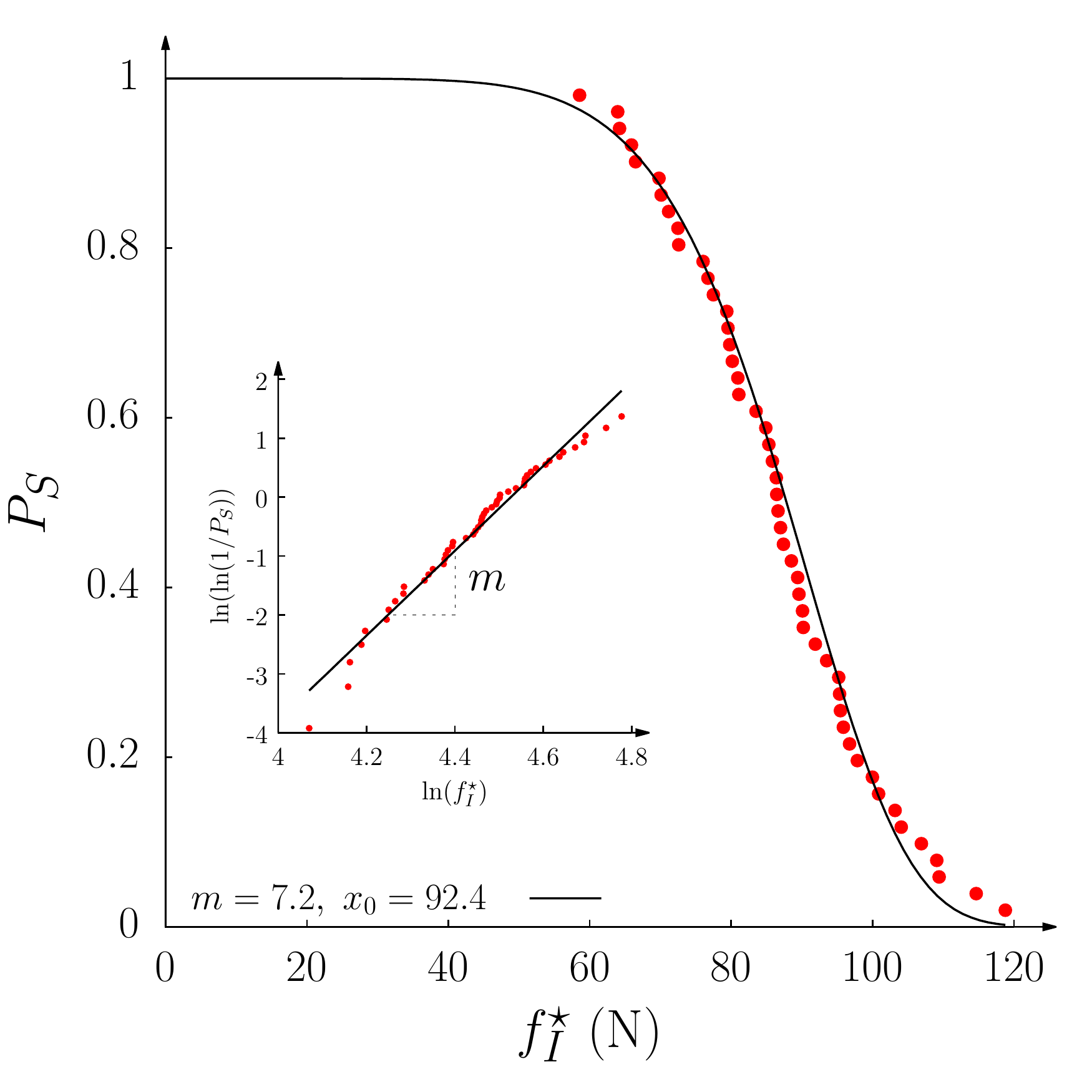}
\caption{Weibull's cumulative distribution function of the tensile yield threshold $f_I^\star$. The link survival is modelled as $P_{S} =1/\exp{({\text{x}}/{x_0})^m}$, where $m$ is the Weibull's modulus, $x_0$ is the scale parameter and $\text{x}=f_I^\star$.
Red points mark data deduced from the experimental campaign, while the curve is a theoretical trend, that was found from fitting equation $\ln(\ln(1/P_{s}))=m\ln{\text{x}}-(m\ln{x_0})$ in a linearised, logarithmic  space (see the inset).}
\label{FigWeibullDetermination}
\end{figure}

A series of strain-controlled Uniaxial Radial Compressions, URC, has been performed to initiate a tensile failure. 
The loading conditions bear a strong resemblance to Brazilian compression test \cite{Carneiro1943}. A ring-cross section of shell was diametrically compressed by the force $F_I$ in which direction the ring deformation $\Delta d$ was measured (Figure~\ref{FigExpNumBT}). A detailed description of the experimental campaign including the characterisation of tested shells can be found in \cite{Stasiak2019}.
Figure~\ref{FigExpNumBT} shows the variety of the mechanical responses captured in the experimental campaign (green curves), but only the average response in terms of maximum force and displacement ($F_I \simeq 122.5\pm19.4$~\text{N}, $\Delta d \simeq 1.83\times10^4\pm0.35\times10^4$~m) was numerically adjusted  (red line in the plot). Since the force law in the links are basically linear, (see Figure~\ref{FigForceLawGlue}a), 
the curve $F_I \leftrightarrow \Delta d$ also obeys the linear relationship with the constant slope $K_I$ -- a shell radial-stiffness that depends on the ring geometry (size, relative thickness, and so forth). Note that also the experimental curves convert to linear trends after a contact stabilisation manifested by the non-linearity of the curve. This stabilisation is caused, in particular, by a surface chipping and/or a slight movement of shell that inhibit the increase of force. The numerical investigations have shown how to model the shell response to URC throughout two micro-parameters: the yield tensile threshold $f_{I}^\star$ and mode-\textit{I} stiffness $k_{I}$. 
The average response presented in Figure~\ref{FigExpNumBT} has been simulated with $f_{I}^\star=85$~N and $k_I=1.1\times10^7$~N/m.
Whereas the link stiffness $k_I$ was found to control $K_I$ varying only the range of $\Delta d$, $F_I$ remained in a linear relationship with $f_{I}^\star$.
Using $f_{I}^\star=F_{I}/\alpha$ with $\alpha=1.414$ (for the average shell with 18~mm diameter and the ring thickness of 2.4~mm), the true \textit{Weibullian} distribution \cite{Weibull1951} of $F_I$ can be easily converted to its numerical substitute shown in Figure~\ref{FigWeibullDetermination}. It characterises the probability of shell survival from the tensile failure with the Weibullian cumulative density function (\textit{cdf}) for scale  and shape parameters, respectively $x_0$ and $m$ \cite{Stasiak2019,stasiak2017}. Using this parameters to distribute $f_I^\star$ on the links in a simulation, the heterogeneity of material, partly caused by geometrical imperfections of shell (with respect to idealised shape of the cluster, Figure~\ref{FigCluster}b), can be introduced.

Finally, we could verify the shape partitioning in view of the breakage context. In case of URC, the cluster model with only circumferential slicing is sufficient to reproduce the experimental breakage into 4 fragments: in the vertical plane recognised as \textit{primary} breakage, and in the horizontal plane as \textit{secondary} breakage (see the insets of Figure~\ref{FigExpNumBT}). However, one must have noticed a distinction in the occurrence of secondary cracks -- $\Delta d$ exceeds the extents observed in the experiments. This is mainly an effect of the model assumptions. More precisely, this modelling did not distinguish the initiation and the propagation of the crack as it arises experimentally -- cracks occurred in just one time step which led to a sudden, high release of energy stored in link (manifested by the drop of the force). After the force mounted up again but in the model we deal with half-rings so the cluster stiffness $K_I$ decreased due to change of geometry.

\begin{figure}[htb]
\begin{center}
\includegraphics[width=0.3\linewidth]{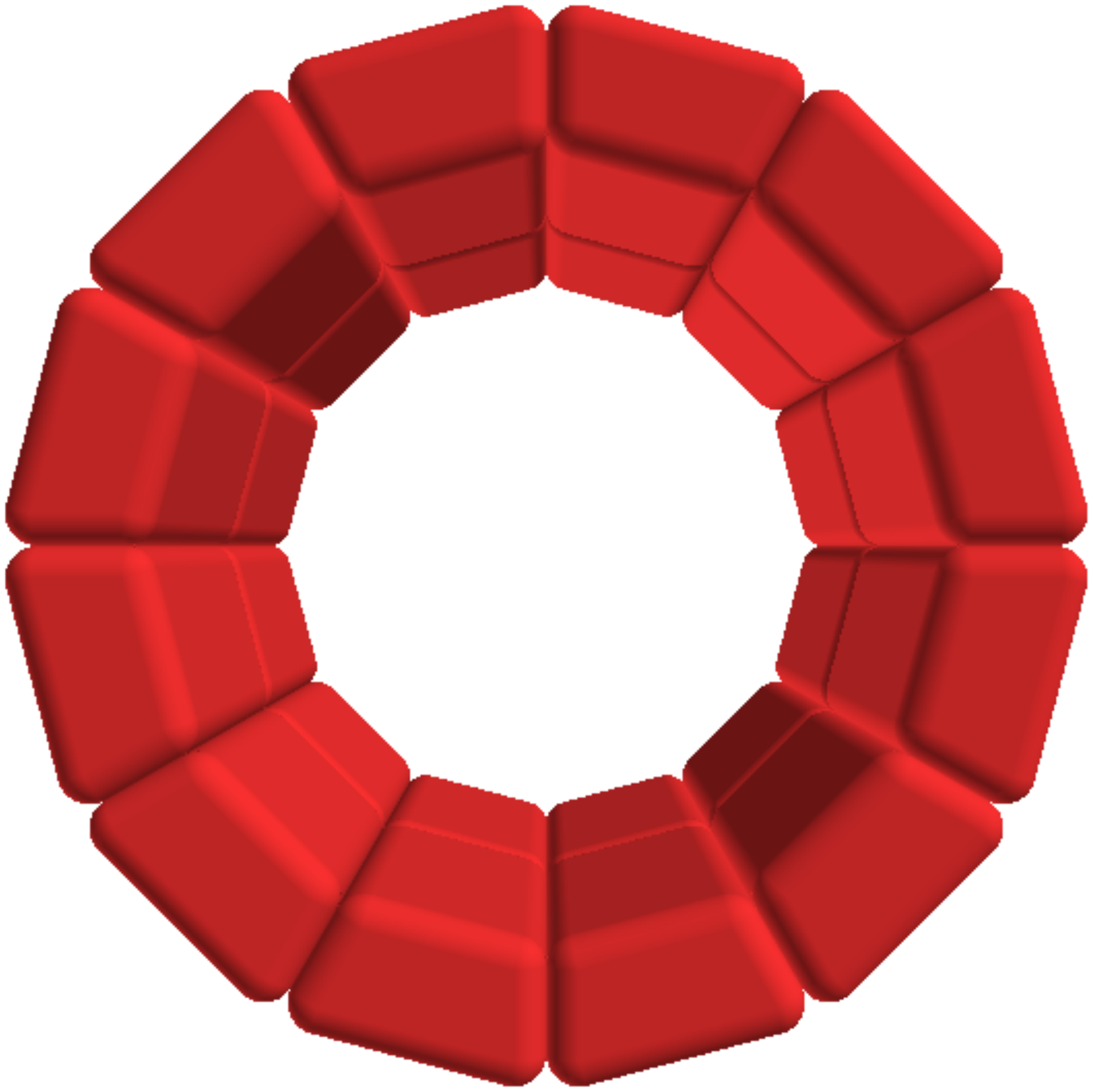}
\includegraphics[width=0.3\linewidth]{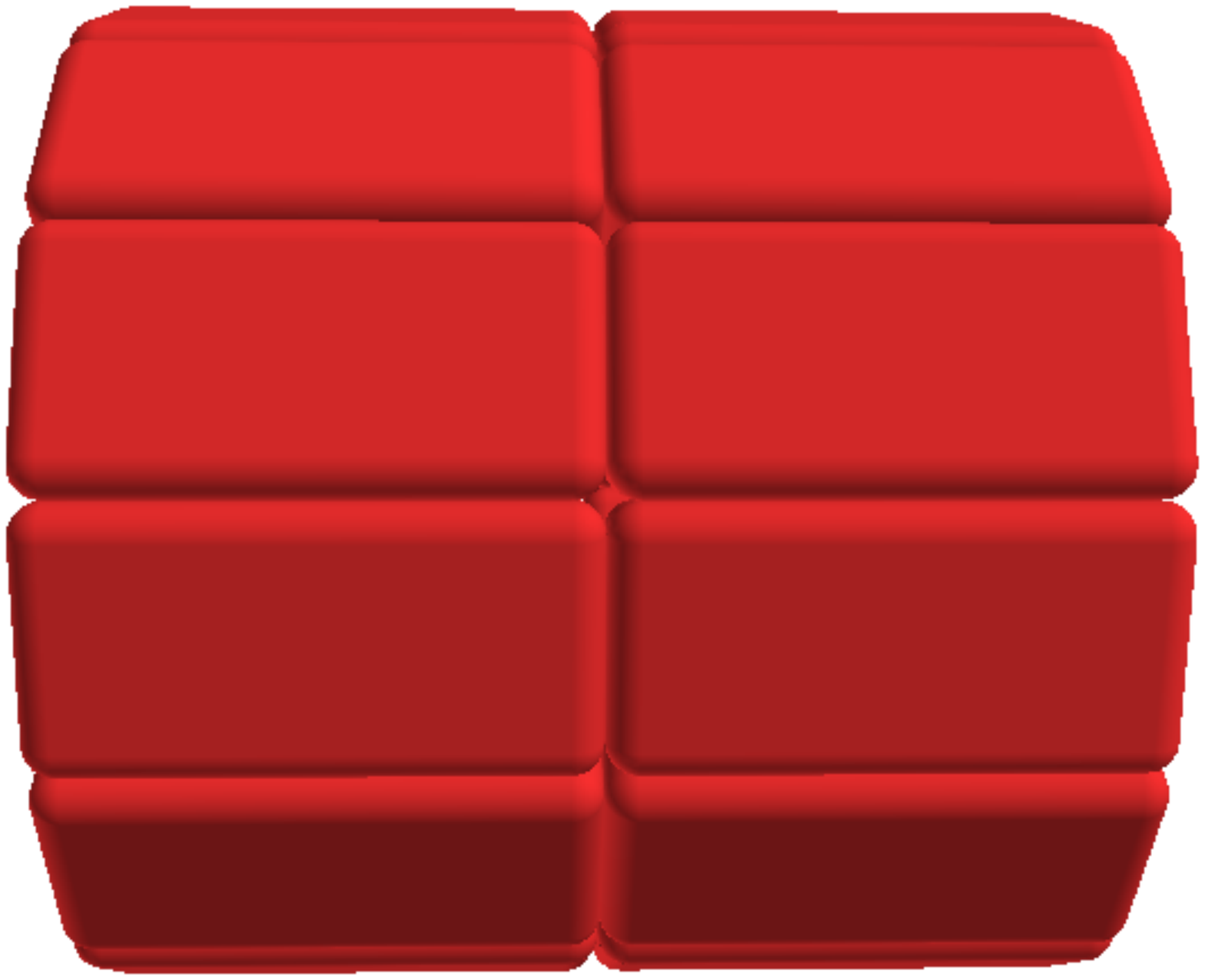}
\caption{Example of a shell partitioned with $N^\star_\text{circ}=12$ and $N^\star_\text{axial}=2$.}
\label{FigShellS2}
\end{center}
\end{figure}

As seen before, the shells are subdivided into $N^\star_\text{circ}$ sectors in its circumference. The sectors are the unbreakable elementary entities of the modelling. In the axial direction, sectors can also be divided into $N^\star_\text{axial}$ parts, Figure~\ref{FigShellS2}.

For oedometer test simulations at low stress levels, that is when the shells tend to fail due to cracks initiated in tension, this shape partition respects the experimentally observed fracture pattern \cite{ReportCermesOedo}. But the higher is the stress level, the more deviate the longitudinal sectors from the actual shape of fragments \footnote{Observations from oedometer tests performed in Laboratory Navier, Paris}. In this work, shell clusters have been sliced by using $N^\star_\text{axial}=1$ and $N^\star_\text{circ}=12$, for most of the cases.  In this way, the compression rate has been consciously limited in the model. Nevertheless, with the prospect of larger scale modelling, reducing the amount of particles in favor of lower computational cost was a necessary initiative. 
Local parameters must be adjusted when the slicing of the shell shape is changed such that: 
\begin{itemize}
\item[\ding{217}] $k_I$ increases proportionally to $N^\star_\text{circ}$ with 24 sectors as a reference 
\item[\ding{217}] $f_{I}^\star$ changes with $N^\star_\text{axial}$ referencing no axial division -- 4 links to break in the axial direction 
\end{itemize}

\subsection{Mode-\textit{II} failure ($k_{II}$, $f_{II}^\star$)}

\begin{figure}[htb]
\centering
\includegraphics[width=0.7\linewidth,]{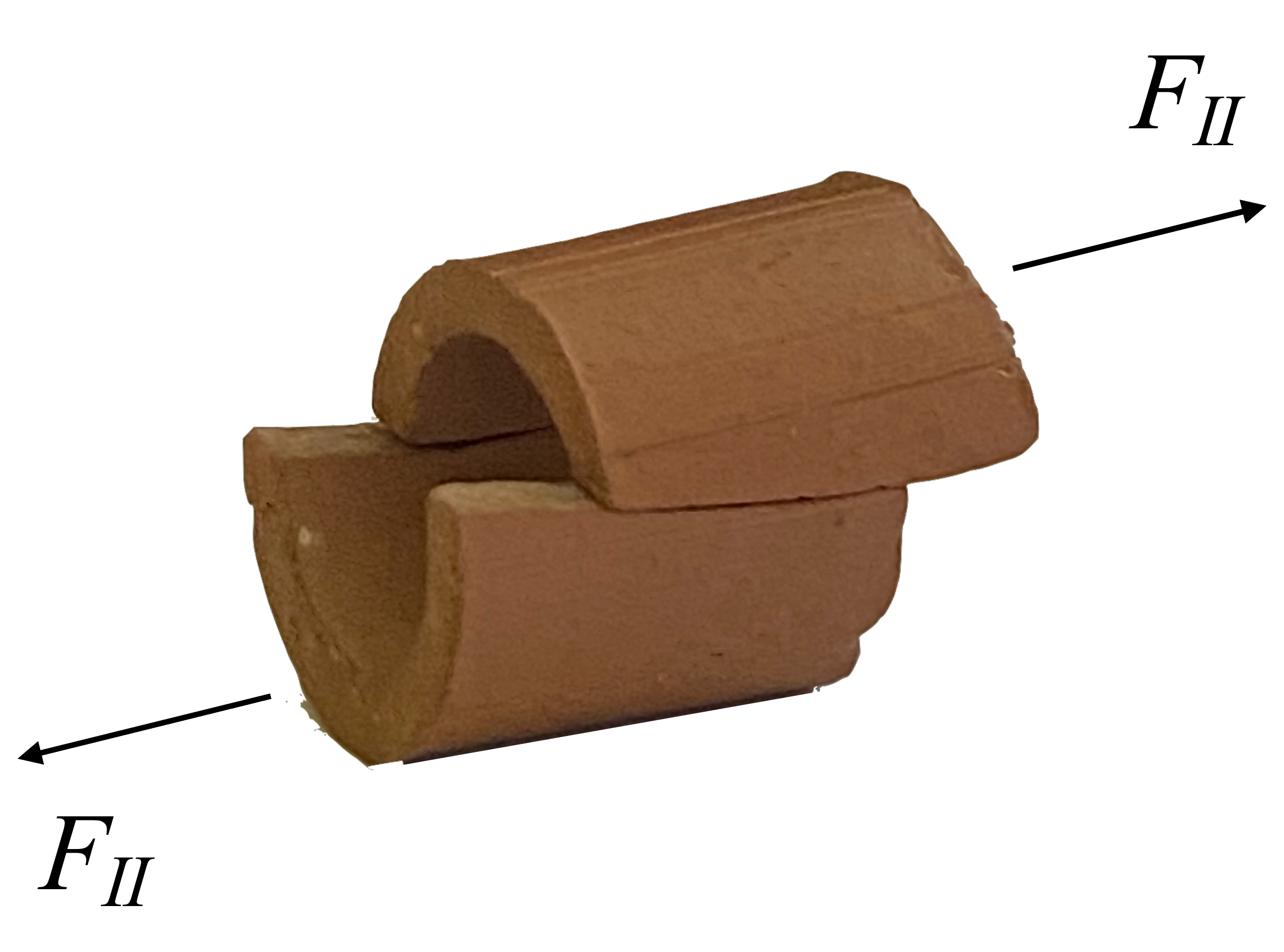}
\caption{The concept of in-plane shear failure. A shearing force $F_{II}$ triggers failure through two axial cracks, a breakage manner directly applied to the radial slicing shown in Figure~\ref{FigCluster}b.}
\label{FigShearScheme}
\end{figure}

In-plane shear tests were performed using an in-house shear device inspired by the commonly used direct shear box such that shell breaks in the middle creating 2 parts; Figure~\ref{FigShearScheme}. The shell is first caged in two-part shear box and then sheared by imposing a constant relative displacement rate. The tangential force was recorded during the test. 
Unfortunately, these attempts of a pure in-plane shear test were hardly satisfying as the shells mainly failed in axial compression for a force  $F_{II} \simeq 2\,000$~N.
As a consequence,  a minimum tangential force of $2\,000$~N was assumed to be needed for mode-\textit{II} failure.

In the current model, four links are set between subsequent sectors. The critical shear strength $f^\star_{II}$ of each link is thus set to 250~N so that the triggers shell-failure in mode~\textit{II}, has illustrated in Figure~\ref{FigShearScheme} is  $F_{II}=2\,000$~N. For sake of simplicity, hereafter the link stiffness $k_{II}$ was set equal to $k_{I}$.

\subsection{Combined mode-\textit{I} and mode-\textit{II} failure ($q$)}

The rupture of a link can result from pure mode-\textit{I} tension or pure mode-\textit{II} shear.
In practice however, the loading state of a link is actually a combination of these two modes, and it is necessary that a failure criterion takes both loading modes into consideration. This is the reason why the yielding function (Equation \eqref{eqYieldFunction}) was introduced (see Figure~\ref{FigForceLawGlue}c).
The yield envelop is cone-shaped when $q=1$, which is similar to a linear Mohr-Coulomb surface with the major difference that the link response remains elastic and linear inside the surface.
Increasing $q$, the yield surface tends to become a prismatic shape, and the combined influence of the two pure rupture tresholds tends to vanish (they become completely independent for $q=\infty$).

A number of URC and in-plane shear simulations, carried out on single shells, have shown a limited influence of $q$ on failure provided that $q \geq 2$.
Finally, a parabolic yield function was used ($q=2$), following \cite{Delenne2002}.

\subsection{Sector-to-sector frictional contact parameters ($k_n$, $k_t$, $\mu$)}

Frictional contact can act between sectors of different clusters, or between sectors of the same cluster if their common adhesion plan has been broken.
One can estimate the contact stiffnesses as follows. {For Young's modulus $E=4$~Ga and the Poisson coefficient $\nu=0.29$, the dimensionless stiffness parameter for 3D Hertz contact expresses \cite{Radjai2011}
\begin{equation}
\kappa = \left( \frac{E}{\sigma_0 (1-\nu^2)} \right)^{2/3} \simeq 267 \quad \textrm{for} \quad \sigma_0=1 \textrm{~MPa}
\end{equation}
where $\sigma_0$ is a reference mean stress. For a typical shell size being the shell diameter $d$ (Figure~\ref{FigCluster}), the normal contact stiffness is $k_n=5.34\times10^6$~N/m, in the case of a linear normal contact law, $\kappa = k_n / (d \sigma_0)$ \cite{Radjai2011}. Recall that we have always been keeping the ratio $k_n/k_t=1$. Then, in case of the cluster with $N_\text{circ}^\star=12$, it was possible to equal the link and frictional contact stiffnesses $k_n=k_t=k_{I}=k_{II}$ at the same value of around $5.5\times10^6$~N/m.}

Concerning the contact friction coefficient, it was estimated using an experimental device of 3 shells described in \cite{Calvetti1997} and \cite{Stasiak2019}. The following value was used: $\mu = 0.24 \pm 0.06$.

There exists other non-numerical parameters that one might find influential for the shell assemblies, \emph{e.g.}, initial density, contact network, orientation of shell. Such investigations cannot be limited to one cluster, and therefore, Section~\ref{secParamStudy} is dedicated to the parametric study at the sample scale, but first a realistic numerical sample needs to be generated. Thus, in the following Section~\ref{secExpeCharacterisation}, we present the experimental identifications of some parameters that serve us as the starting points.

\section{Microscale characteristics of a real shell assembly\label{secExpeCharacterisation}} 

\begin{figure*}[htb]
\centering
\includegraphics[width=0.35\linewidth, trim={2cm 2.5cm 2cm 3.5cm},clip]{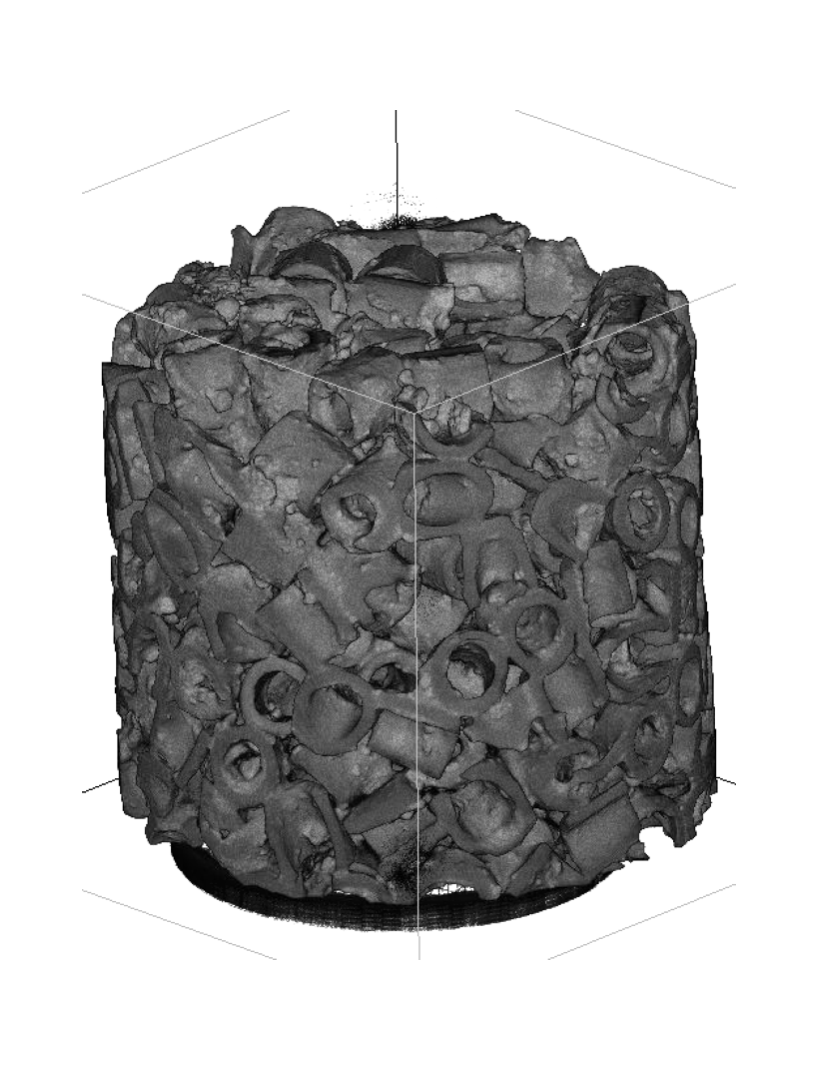}\hspace{0.1cm}
\includegraphics[width=0.63\linewidth, trim={0cm 0cm 1.5cm 0cm},clip]{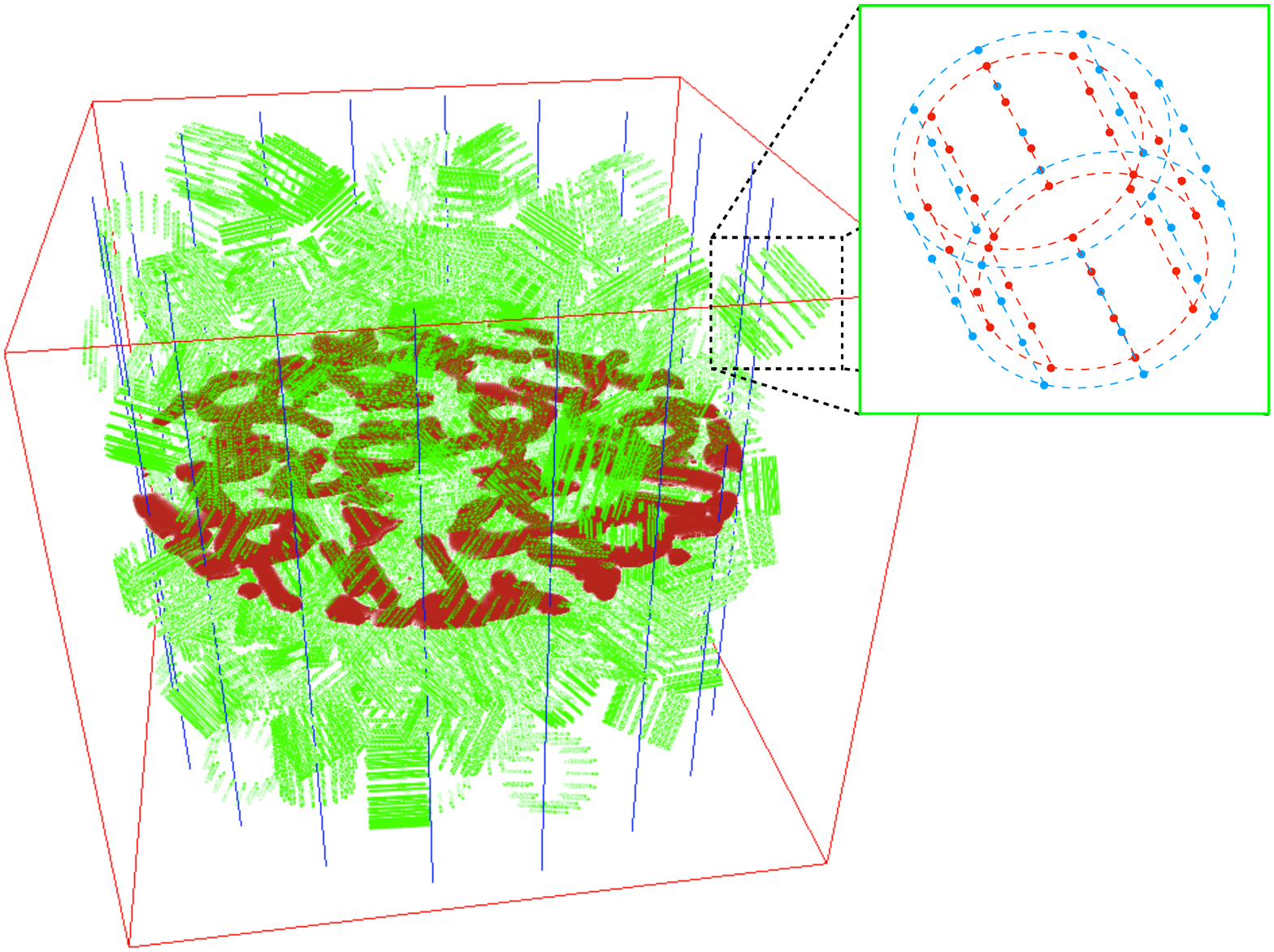}
\begin{flushleft} \hspace{2.8cm} {\large(a)} \hspace{7.5cm} {\large(b)}  \end{flushleft}
\caption{X-ray tomography: \textbf{9a} -- A reconstructed volume of a true sample obtained with X-Act software based on the Filtered Backprojection (FBP) algorithm.
\textbf{9b} -- Shells detection with \texttt{3DShellFinder} tool, for the same sample. The red zone represents the solid phase of a horizontal slice cut out from the volume (\textbf{9a}). Each successfully found shell is represented by a number of the search points (see inset), and those points are marked in green. Whereas red box marks the border of full size image, the blue vertical lines show the reduced zone within which the algorithm detects the shells.}
\label{FigX-rayVolume}
\end{figure*}

Let us start with the internal variables that on various complexity level refer to the fabric of granular material. Probably, the most basic measure is a density. This study relay on the number density $n$ measuring the amount of shells in the volume unit. The true values that the modelling aims are $n=151\,384\pm1\,319~\text{m}^{-3}$ for the sample with 2.09~$\text{m}^{3}$ volume and $n=155\,129\pm3\,952~\text{m}^{-3}$ for the volume of 6.43~$\text{m}^{3}$. The given values concern the samples being a shell assembly, a situation we will address numerically (Section~\ref{secParamStudy}).

\begin{table}[htb]
\centering
\begin{tabular}{lr}
\toprule
 Setting & Value \\
\midrule
Voltage                & 135~kV \\
Current                & 500~$\mu$A \\
{Maximum spatial resolution}     & 100~$\mu$m\\
Spot size              & large \\
No. angular positions  & 1\,440 \\ 
No. scans per position & 6 \\
{Highest frame rate}             & 5 \\
\bottomrule
\end{tabular}
\caption{{The image acquisition capabilities of the RX-Solutions scanner} with an ``indirect'' flat-panel Varian PaxScan\textsuperscript{\textregistered} 2520V detector and Hamamatsu Corporation L8121-03 source \cite{Viggiani2015}.}
\label{TabXray}
\end{table}

Towards more comprehensive characterisation, a couple of cylindrical samples with diameter $\simeq 12$~cm and height of about the same size were extracted from the compressible layer of the tunnel segment, VMC. On that account they are the shell-mortar composites. Thanks to X-ray radiographs acquired in RX-Solutions scanner of Laboratoire 3SR, it was possible to reconstruct their volumes (Figure~\ref{FigX-rayVolume}a). {The acquisition capabilities of the RX-Solutions scanner} are specified in Table~\ref{TabXray}, but the precise descriptions of the X-ray imagining technique and the scanner itself are not the object of the current paper, yet can be easily found, for example, in \cite{Viggiani2015}. 

\subsection{Shell detection in 3D images}

Form a binary image, that is a 3D solid-void matrix, we can extract a couple of datasets.
A spacial detection of the shells motivated us to perform the X-ray scanning in the first place. The full detection concerns two components: the position and the orientation. 
\texttt{3DShellFinder} algorithm was specifically designed to suit the tubular geometry of the shell. A perfect tube is pre-specified such that it can be inscribed in the true body of the shell. This tube is then represented by a set of the search points as shown in the inset of Figure~\ref{FigX-rayVolume}b. 
For each targeted shell, the algorithm probes different positions and orientations varying them by a considerably large amount of the small increments. An attempt to detect shell firstly consists in assigning the search points to the voxels in the image. Their grey level equals either 1 for voids or 0 for solid. Then, the algorithm proceed with the minimisation of an error function:
\begin{equation}
\label{EqError}
\mathcal{E}(\bm{x}_\text{shell},\ \hat{\bm{q}}_\text{shell}) = 1 - \frac{1}{N_\text{ZOI}} \sum_{i \in \text{ZOI}} \mathbb{I}\left[\bm{P}_i(\vec{x}_\text{shell},\ \hat{\bm{q}}_\text{shell}) \right]
\end{equation}
where $\mathbb{I}$ is the 3D-image and $N_\text{ZOI}$ is the number of search points that are located within the zone of interest (ZOI). The function of Equation~\eqref{EqError} is parametrised by the position (vector $\bm{x}_{shell}$) and the orientation (quaternion $\hat{\bm{q}}_\text{shell}$) of the search points $\vec{P}_i$.
For the search points perfectly aligned with the shell in the image $\mathcal{E}(\bm{x}_\text{shell},\ \hat{\bm{q}}_\text{shell}) =0$, but to increase the probability of the successful search the algorithm accepts a small error, \textit{e.g.}, in the following case $\mathcal{E}(\bm{x}_\text{shell},\ \hat{\bm{q}}_\text{shell})\leq0.02$.
For the sample of Figure~\ref{FigX-rayVolume}a, the search tube was circumferentially discretised into 18 points for two rings of radii 70 and 80~voxels. Axially, 30 points were densely distributed along the length of 120~voxels. 
Figure~\ref{FigX-rayVolume}b shows 229 shells (green points) successfully located in the image.
We have estimated\footnote{From the experimental measurement of $n$.} that the algorithm identified around 98~\% of shells (positions and 3D orientations). 
The unsuccessful searches were caused by (\textit{i}) the damaged of the boundary shells or (\textit{ii}) a strong imperfection of the full shell within the sample core.

\subsection{An approach to analyse shell orientations}

\begin{figure}[htb]
\centering
\includegraphics[width=0.98\linewidth,trim={1cm 0 7cm 0},clip]{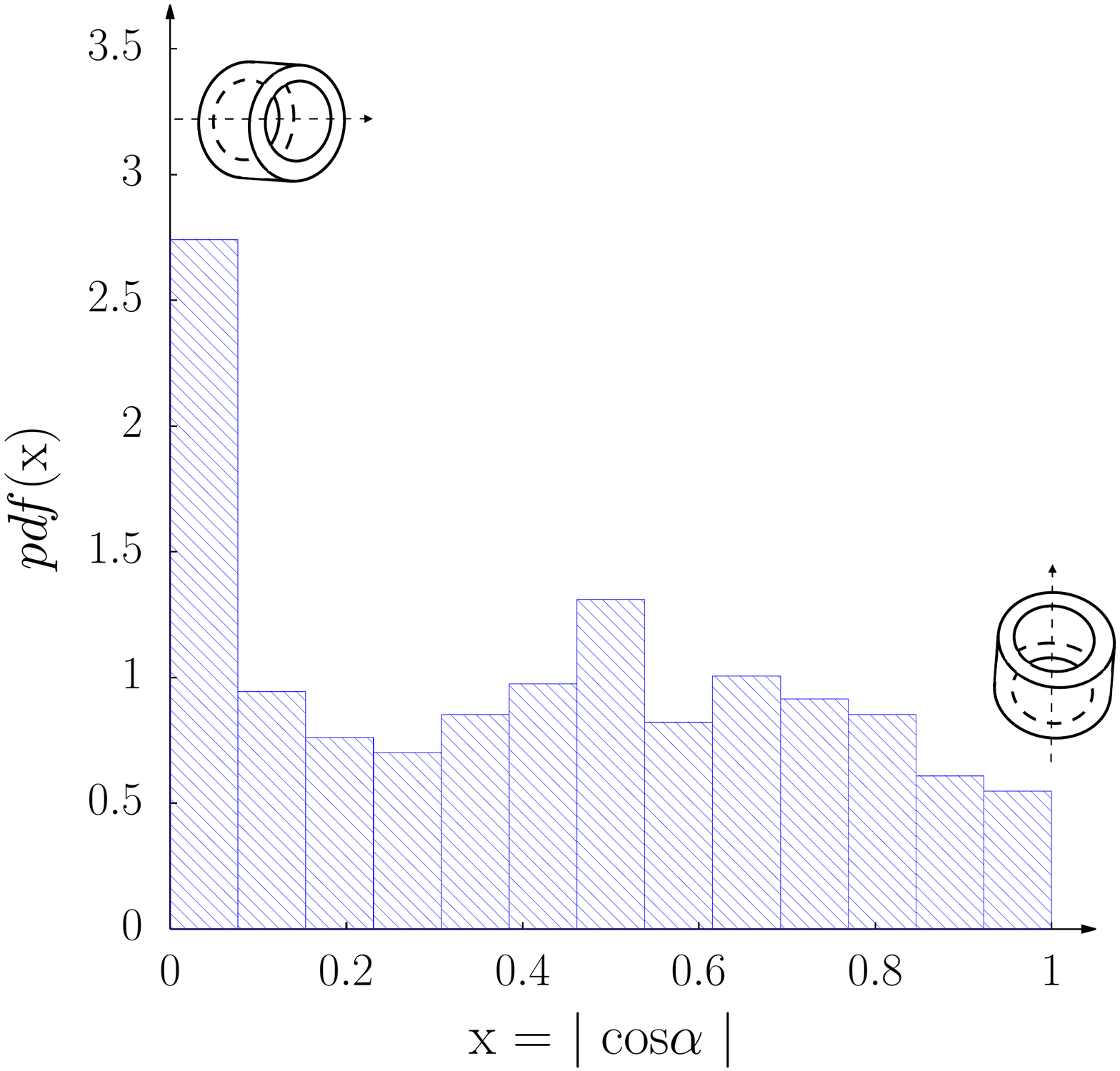} 
\caption{Probability density function \textit{pdf(\emph{x})} of shells orientation from all the samples cut out of a compressible layer around a tunnel segment (\textit {VMC}). $|\cos\alpha|$ varies between 0 for horizontal orientation and 1 for vertical shells.}
\label{FigExpOrient}
\end{figure}

More complex aspect of studying the sample fabric should include a combined treatment of the particle orientation and the contact network \cite{Oda1972}. Experimentally, with the X-ray radiographs we were able to address only the first one. 
Considering the axisymmetric shell, the ring cross-section defines the weak plane, while a central axis indicates its strong direction. It is worth mentioning that in the global coordinate system the vertical axis corresponds to the loading direction.  Hence, an investigation of the shell/cluster orientation can be limited to the inclination of the central shell axis with respect to the vertical. This inclination can be measured by an angle $\alpha$, but herein it is quantified indirectly with its cosine, such that $|\cos\alpha|=1$ for vertically resting shells and $|\cos\alpha|=0$
for the horizontal orientation as in URC tests. Figure~\ref{FigExpOrient} presents the probability density function of the shell orientation with a strong but noncontinuous anisotropy proven by a well marked peak for $|\cos\alpha|=0$. Despite a natural tendency of the shell to embed horizontally, there still exists a partial isotropy for the remaining orientations. As a consequence, in oedometer test the shells are mainly radially compressed, and one can expect mainly the tensile stress to cause the primary breakage.

Additionally, a void-to-solid ratio $e$ could be easily extracted the from 3D image, where solid includes both the shells and the mortar. For the sample shown in Figure~\ref{FigX-rayVolume}a, the standard void ratio $e$ equals to 1.283 for both shells and mortar. The calculation disregards the boundaries of sample that were spoiled due to a extraction process used to drill the specimen from the tunnel segment. 
A rough estimation  of $e$ for shells (without taking into account the mortar) only reaches 2.4. A difference between those values will play a crucial role in the evaluation of the shell-mortar modelling where the volume mortar will not be represented (not further discussed in current paper). 

\section{Sensitivity to the model inputs}
\label{secParamStudy}

As seen in the Section~\ref{secPramMicro}, the numerical model requires numerical, mechanical and geometrical parameters. In order to carry out this study, it was fundamental to first perform a sensitivity analysis and to try to really understand the effects of the parameters.

The oedometer test consists in imposing an uni-axial compression while the boundaries are fixed in remaining directions (Figure~\ref{FigExpMechBeh}).
The mechanical response is displayed as the relationship between the axial stress $\sigma_a$ and the natural axial strain:
\begin{equation}
\label{EqStrain}
    \varepsilon_a = \ln(H/H_0)
\end{equation}
where $H$ and $H_0$ are the current and initial high of the sample, respectively.

The current analysis is essentially a bare judgement of the consequences of the parameter variations on the macroscopic mechanical response. The present section attempts to make clear which are the most influential parameters and how through their value control the mechanical response.

To generate a specimen of packed clusters, a deposit under normal gravity have been simulated.
Initially, the clusters were placed, without contact, within a cylindrical container closed by a flat lid on top; the threshold forces  of the bonds that hold the cluster segments were set so that the clusters can not break in this phase. Each cluster has been not only positioned in a separate node of a cylindrical grid but also randomly oriented.
Afterwards, the suspension deposited due to the activation of gravity. Additionally, an off-grid movement has been imposed at each cluster. Such numerical trick is equivalent to the sample shaking which prevents formation of local cavities, especially within highly frictional assemblies. Once the assembly embedded at the bottom of the container, the energy dissipation process needed time to reach a satisfactory equilibrium state.

While falling, the clusters collide through frictional contacts (see Figure~\ref{FigForceLawGlue}b, dashed line). The greater is the friction coefficient, the looser is the cluster packing at the end.
With series of numerical deposits, an analytical relationship between the inter-granular friction coefficient $\mu_{shell}$ and the density has been determined \cite{Stasiak2019}. Therefore, it was possible to tune the initial density. 
The deposit was simulated with $\mu_{shell}$ set to 0.08, with friction-less walls ($\mu_{wall}=0$). The sample density is then in the order of the experimental one {(see section  \ref{secExpeCharacterisation})}. 

{
In the following, the sensitivity to different parameters is discussed in the context of oedometric compressions of cluster samples, at constant loading rate. The loading rate is chosen low enough to ensure a quasi-static regime.  To do so, we ensured that the inertial number $I = \dot{\varepsilon}_a \sqrt{m / \sigma}$ was less than $10^{-3}$ \cite{Combe03,GDRMidi}. Additionnaly, the time step was chosen to ensure the numerical stability of the integration scheme for the equations of motion. 
}

\subsection{Mechanical and numerical interaction parameters} \label{secMicroMechParam}

\subsubsection{Parameters of non-bonded frictional contacts}

\begin{figure}[htb]
\centering
\includegraphics[width=0.98\linewidth,trim={20 0 0 20},clip]{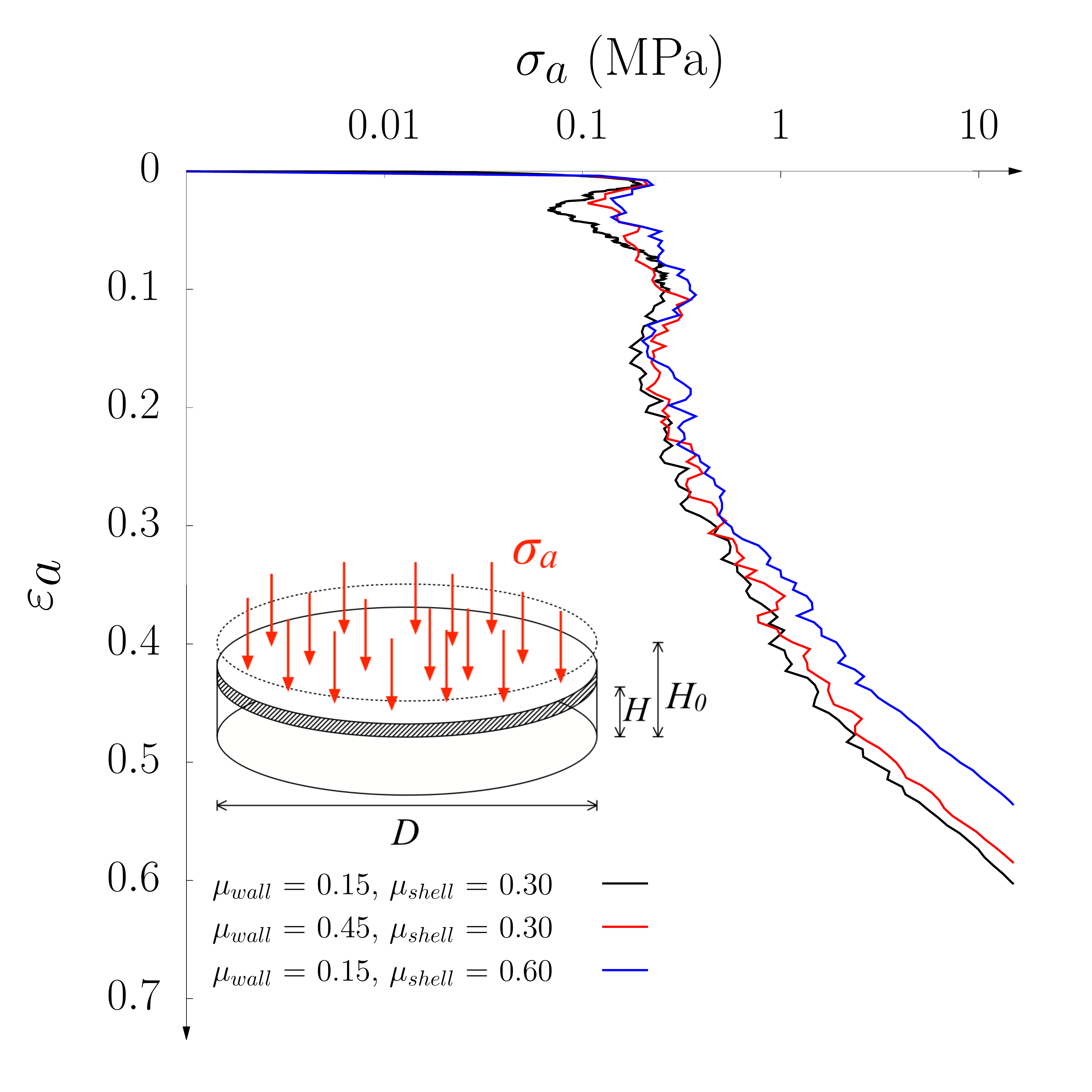}
\caption{Oedometeric curves for various value of friction coefficient between clusters $\mu_{shell}$ and between clusters and walls $\mu_{wall}$. Inset: Loading condition.}
\label{FigFrictStrStr}
\end{figure}

Figure \ref{FigFrictStrStr} shows stress-strain trends obtained numerically for a dense sample for selected couple of friction coefficients, $\mu_{shell}$ or $\mu_{wall}$.
No clear influence of any friction coefficient can be noticed. It can however be noticed that for a given strain, the vertical stress increases with the friction coefficients.
At the initial state, before loading, nearly 89\% of the interactions are bonded and 11\% are frictional contacts. During the compression, some links break and the later percentage increases.
However, an oedometer compression, unlike classical shear loading, does not massively mobilize friction between the constituent rigid elements. This can explain the quite minor influence of the frictional parameters on the compression curves.

The other parameters of non-bonded contacts are the stiffnesses. To avoid unreasonable overlaps, $k_n$ and $k_t$ were varied in the order of $10^6$ to $10^7$~N/m. In considered range, their values were not observed to significantly effect the macroscopic response.

\subsubsection{Cohesive links}

\begin{figure}[htb]
\centering
\includegraphics[width=0.98\linewidth,trim={0 0 0 0},clip]{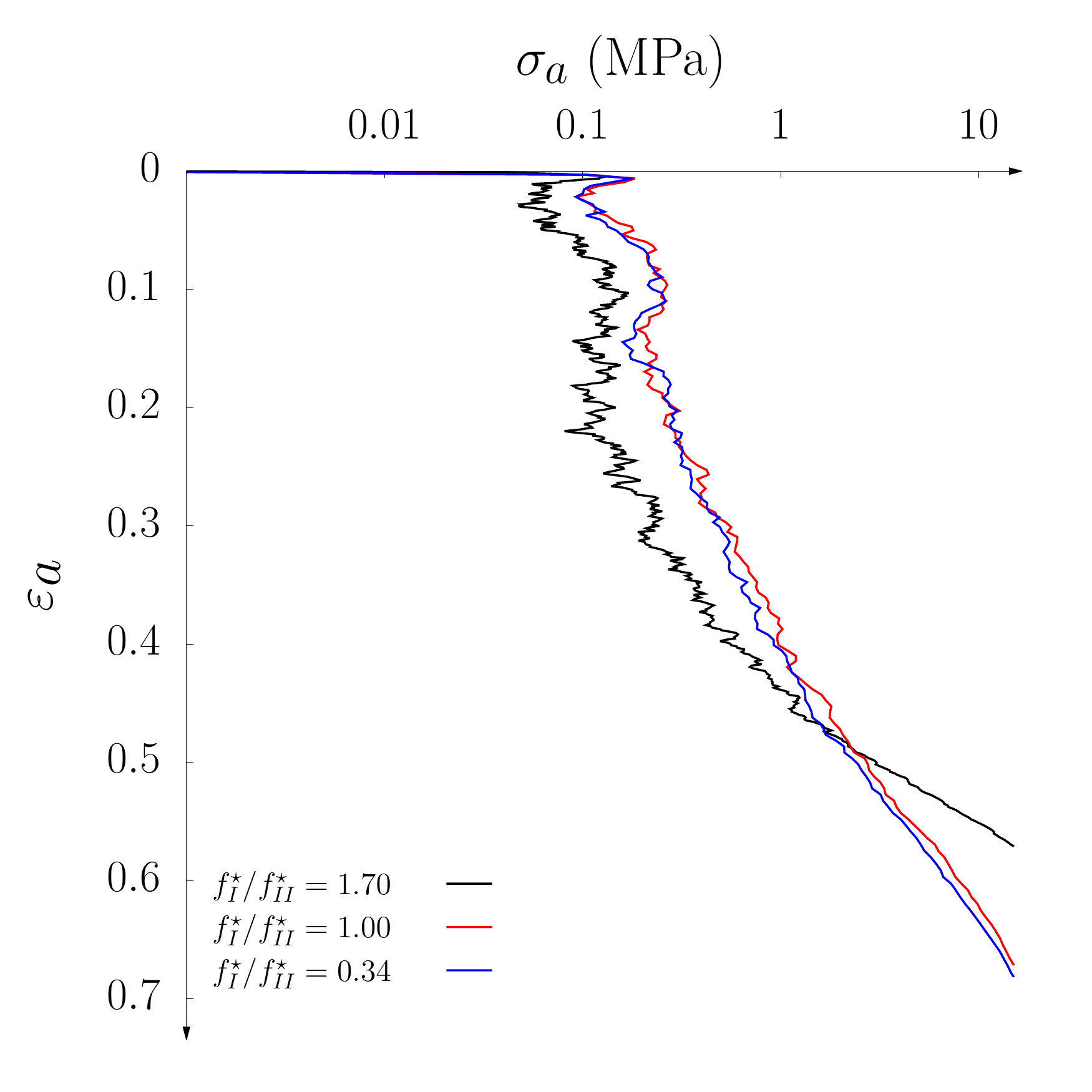}
\caption{Oedometrics curves for various values of the tensile over shear yield limit ratio $f_{I}^\star/f_{II}^\star$.}
\label{FigShearForStrStr}
\end{figure}

Analogous arguments must be made concerning the stiffness of the link, both in mode-\textit{I} and mode-\textit{II} direction. Particularly, $k_I$ was well estimated from the experiments, %
and it does not seem meaningful to study the sensitivity to this parameter.
The key to the understanding of the mechanical response to oedometric compression is the strength of clusters, controlled through the yield thresholds (see Section~\ref{secModel}),  $f_{I}^\star$ and $f_{II}^\star$.
Whereas tensile yield limit $f_{I}^\star$ has been kept constant over simulations the shear one $f_{II}^\star$ was varied, $f_{I}^\star/f_{II}^\star = 0.34,\ 1.00,\ 1.70$. Figure~\ref{FigShearForStrStr} makes clear that the condition $f_{II}^\star \ge f_{I}^\star$ is sufficient to disable the influence of $f_{II}^\star$. For $f_{I}^\star/ f_{II}^\star = 1.70$, a remarkable influence of the shear threshold exists. It results in a decrease of the plateau and a stiffening appearing at a lower axial strain.

\begin{figure}[htb]
\centering
\includegraphics[width=0.98\linewidth]{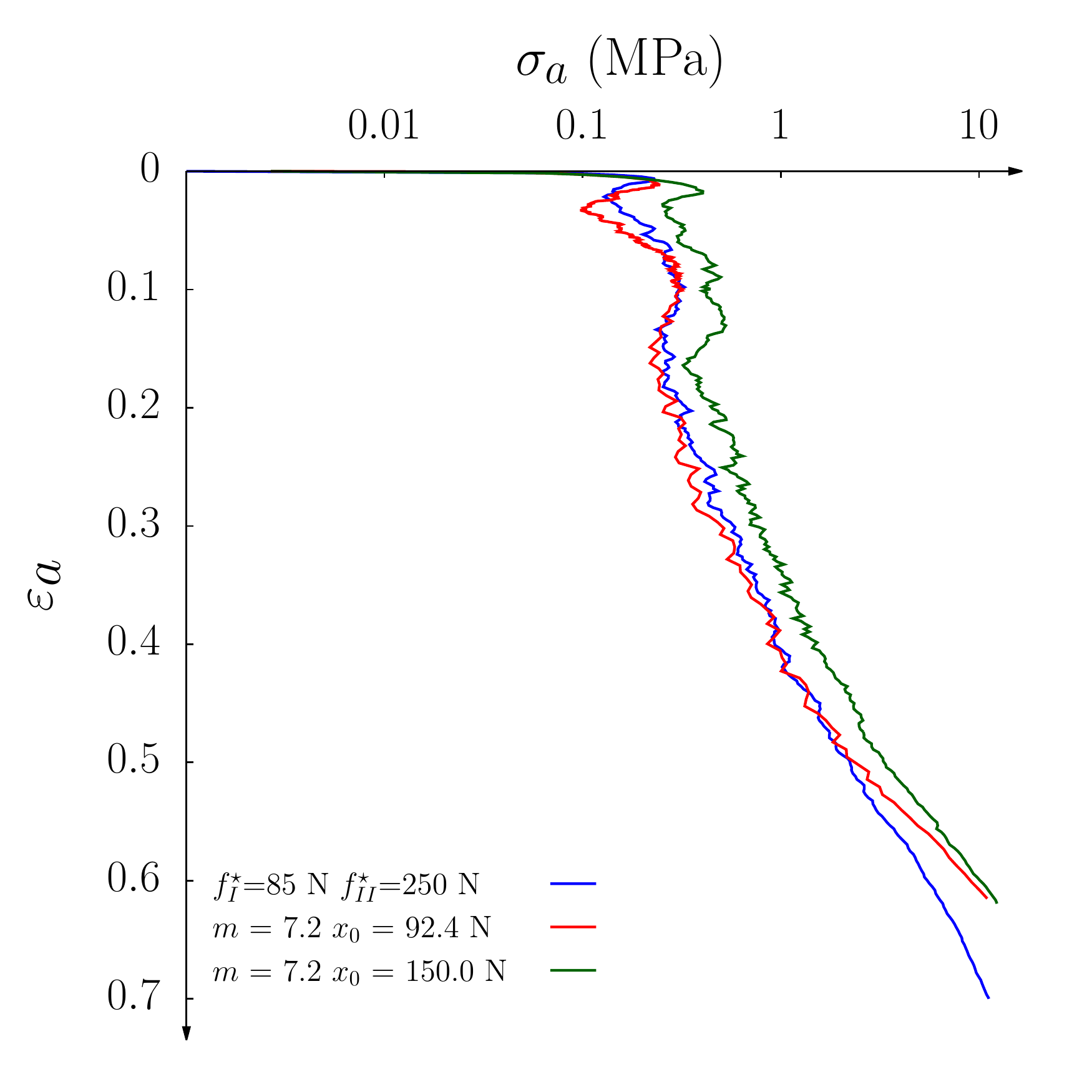}
\caption{Influence of the tensile strength on the macroscopic response to uniaxial compression. Modelling with a strength variability according to Weibull distribution (Figure~ \ref{FigWeibullDetermination}).}
\label{FigtensileThresholdWeibullStressStrain}
\end{figure}

As discussed in Section~\ref{secPramMicro}, the shells exhibits strong variability of the tensile strength characterised by the weibullian distribution; Figure~\ref{FigWeibullDetermination}. In such a distribution \cite{Weibull1951}, the probability density function is described as a combination of power law and exponential function:
\begin{equation}
P_S = \left( \exp (x/x_0) \right)^{-m}
\end{equation}
It is ruled by the modulus or shape parameter $m$, and the scale parameter $x_0$. A modulus $m<10$ corresponds to a wide distribution, \emph{i.e.}, a strong variability. 
Here, instead of using a constant value of $f_I^\star$, we choose to take into account a slight variability whose statistical distribution follows a Weibull distribution whose parameters are determined experimentally: $m=7.2$, $x_0=92.4$~N (see Figure~\ref{FigWeibullDetermination}). Since $f_{II}^\star=f_I^\star/0.34$ for each cluster the variability of cluster strength apply both on $f_{I}^\star$ and $f_{II}^\star$. 
Figure~\ref{FigtensileThresholdWeibullStressStrain} presents the macroscopic stress-strain response with and without weibullian variability and explores the increase of $x_0$ (that is an assembly of stronger and stronger shells having the same breakage variability). 
The modelling with no variability employs only the average strength ($f_I^\star=85$~N and $f_{II}^\star=250$~N).
It is shown that the variability has a minor influence on the mechanical behaviour. The sample structure already provides a diversity in local fabrics and thus on the individual cluster failures. Thus, we state that an homogeneous cluster strength is sufficient in our simulations ($f_{II}^\star$ and $f_I^\star$ have constant values). On contrary, the level of this strength plays a key role. The higher is the tensile yield threshold $f_I^\star$ (also represented by $x_0$), the higher axial stress withstands the sample. 

\subsection{Internal structure influence} \label{secInitialParam}

\subsubsection{Initial density of the packing}

\begin{figure}[htb]
\centering
\includegraphics[width=0.98\linewidth,trim={0 0 0cm 0},clip]{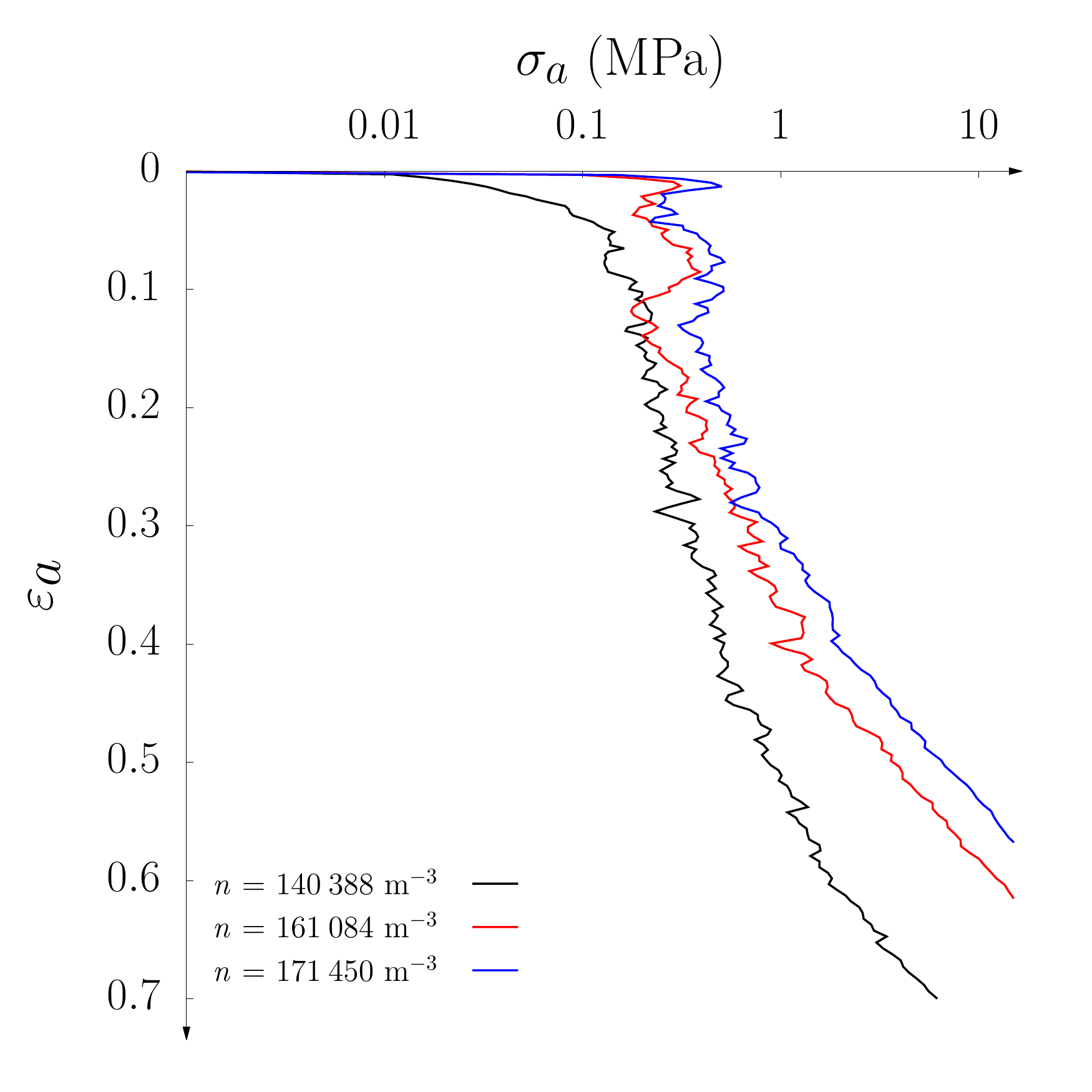}
\caption{Influence of the initial density (number of clusters per $m^3$) of the specimen for uniaxial compression.}
\label{FigDensity}
\end{figure}

Initially, the numerical campaign focused on dense assemblies whose initial density was of the same order of magnitude as the experimental measurements. 
Here, it is proposed to account for the influence of the initial density of the samples on the oedometer test response; the density being controlled by the number of particles per unit volume, denoted $n$. Three densities were tested. The mechanical response curves are given in the Figure~\ref{FigDensity}.

For very small strains, $\varepsilon_a<0.01$, the three curves appear superimposed, revealing equivalent initial stiffness. Once the stress reaches 0.01~MPa, the 3 curves differ, with the loosest specimen (black curve on the Figure~\ref{FigDensity}) showing a slow hardening phase while the other two denser specimens see their strengths take off towards a peak at 0.3~MPa for the red curve (intermediate $n$) and almost at 0.7~MPa for the densest specimen (blue curve). This result is finally quite classical. It is comparable to what can be observed on standard granular materials: the higher the density, the higher the average number of contacts per grain and the higher the strength of the material for low strains. 
For $0.05<\varepsilon_a<0.5$, the intermediate $\sigma_a$-plateau has the highest slope for the lowest cluster density specimen, with all clusters being broken around $\varepsilon_a=0.5$. For the two other densest specimens, the $\sigma_a$-plateaus have similar slopes and end for  $\varepsilon_a=0.4$. Once this intermediate regime is over, the hardening phase begins (``linear'' relationship in a semi-logarithmic scale). The initial density seems to play only a secondary role here, all the clusters having been broken.

On one side, Figure~\ref{FigDensity} makes clear that the lower is the initial density, the larger is the final strain range. On the other side, the stress level is significantly lower despite the identical cluster strength.

Against our expectations, the loosest specimen exhibits the oedometer behaviour closest to the experiments whereas it has a high density of shells. When the DEM specimens have an initial cluster density similar to that observed experimentally with shells, the oedometer behaviour differs from the experiments in that the initial stiffness of DEM specimen is higher. 

A possible interpretation is that, for the same density, the DEM specimens have a higher coordination number than the experimental samples. This interpretation is motivated by the fact that the shells are of rather irregular shape, which is not the case of our digital clusters which then tend to organize themselves spatially and thus have a greater number of contacts.  

\subsubsection{Anisotropy of cluster orientations} 

\begin{figure}[htb]
\centering
\includegraphics[width=0.98\linewidth]{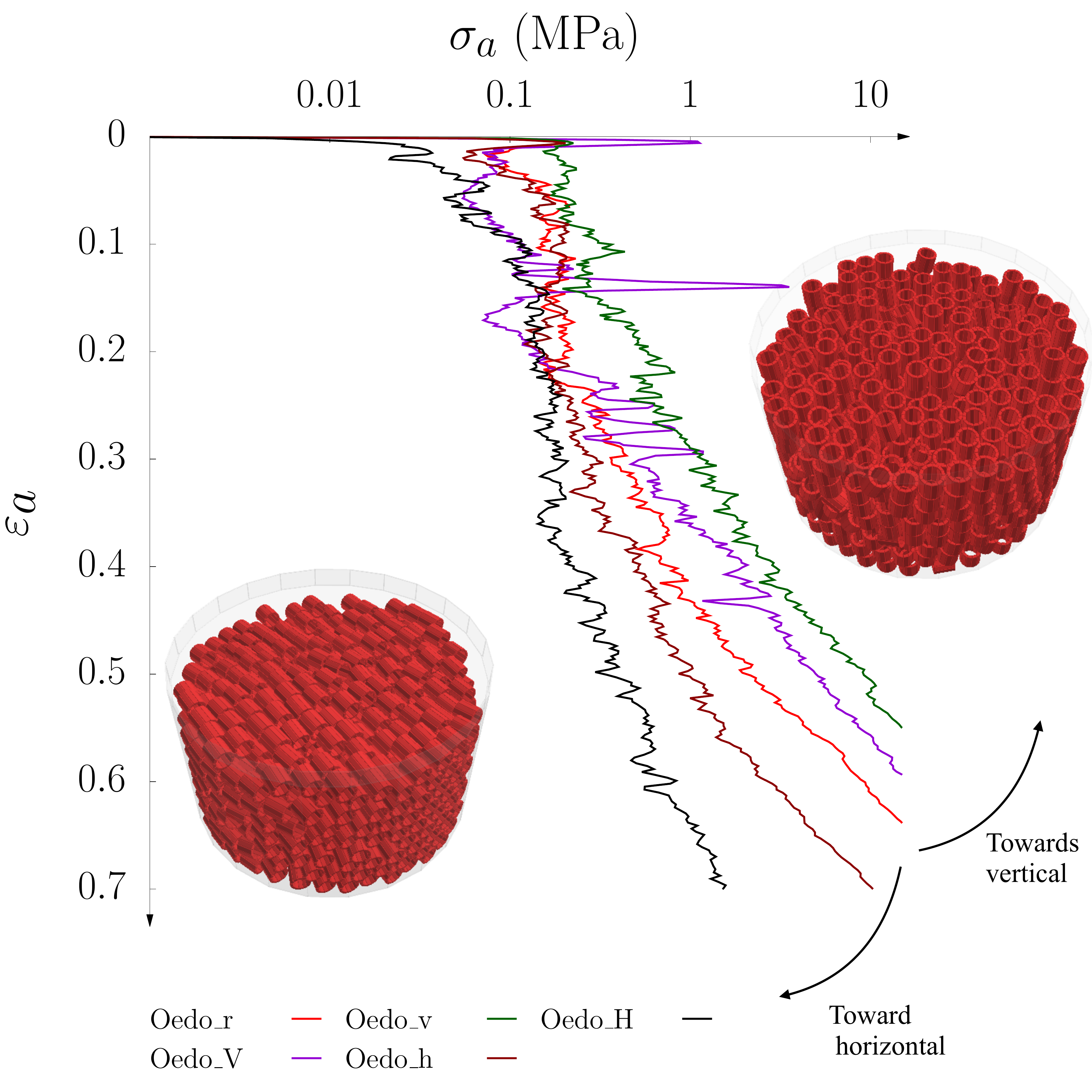}
\caption{Influence of the cluster orientations on the macroscopic response to uniaxial compression. Five different specimens with different cluster orientations. The axis orientations of the clusters in the packing before being compressed are: \emph{Oedo.V} (all vertical), \emph{Oedo.v} (many vertical); \emph{Oedo.r} (randomly oriented), \emph{Oedo.H} (all horizontal), \emph{Oedo.h} (many horizontal).}
\label{FigTensileForWeibStrStr}
\end{figure}

In this section, we focus on the influence of another very influential parameter which is the cluster orientations in the initial packings. Indeed, as for clay shells, the maximum cluster strength depends on the loading direction; the maximum strength in mode-\emph{I} being much weaker than in mode-\emph{II}, as seen in the Section~\ref{secPramMicro}.
Therefore, five different specimens were prepared using a specific numerical procedure (not explained in this paper) to control the orientation of the clusters. For specimens \emph{Oedo.H} and \emph{Oedo.V}, the clusters were \underline{all} oriented horizontally and vertically, respectively, with a given tolerance. An illustration of these two specimens can be seen in the Figure~\ref{FigTensileForWeibStrStr}. For specimens \emph{Oedo.h} and \emph{Oedo.v}, the clusters are mainly (but not only) oriented horizontally and vertically, respectively. Finally, for specimen \emph{Oedo.r}, the clusters are randomly oriented, like it is on the real samples, see Figure~\ref{FigX-rayVolume}.

Figure~\ref{FigX-rayVolume} shows the mechanical responses to uniaxial loading for these five specific specimens. 
As anticipated, vertically oriented particles tend to improve compressive strength; it is perhaps more appropriate to say that the greater the number of horizontal particles, the weaker the compressive strength.
The specimen \emph{Oedo.r} (red curve) shows an intermediate strength.

\subsection{Comparison of the grading curves throughout the shell-breaking process}

\begin{figure*}[htb]
\centering
\includegraphics[width=0.45\linewidth]{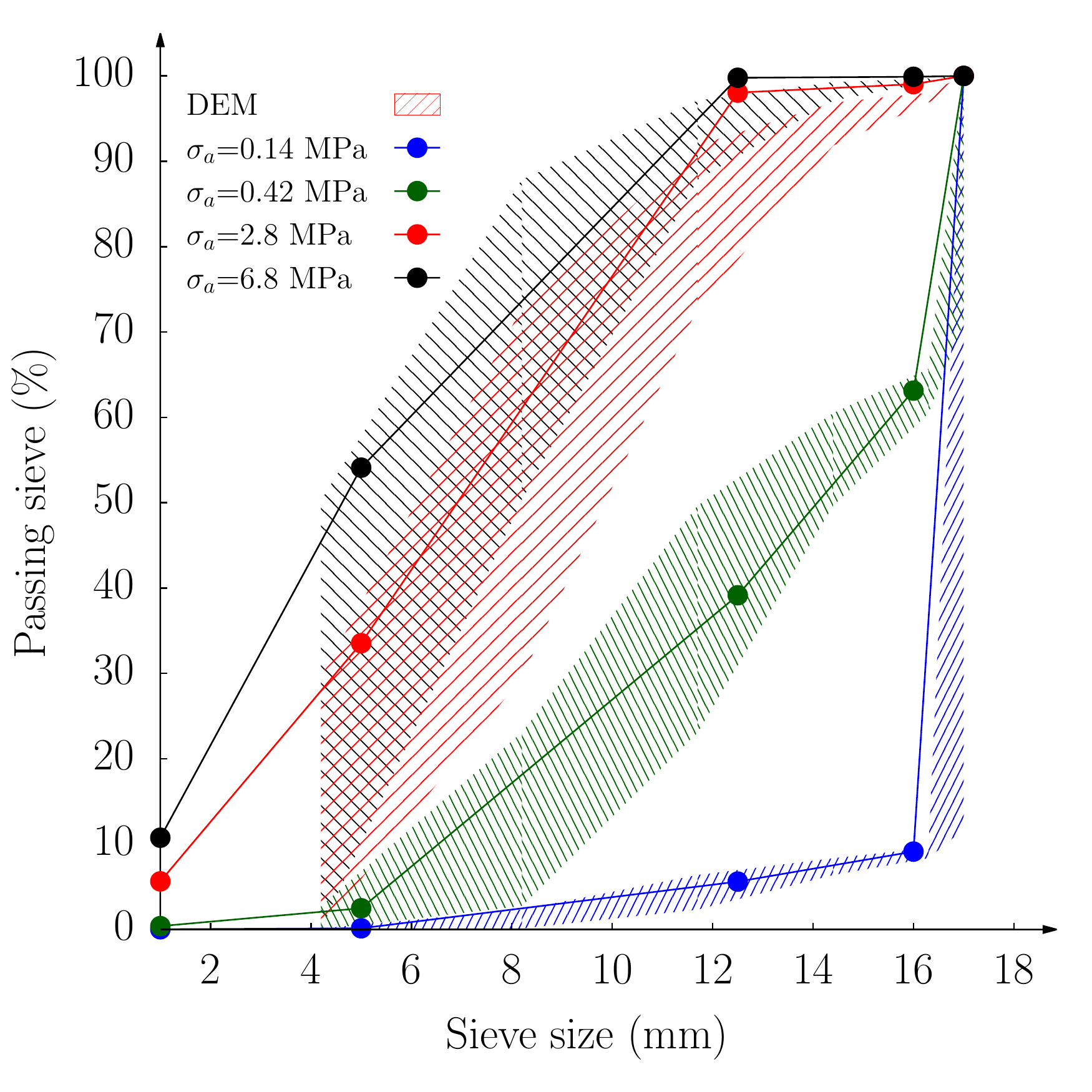}\hspace{0.5cm}
\includegraphics[width=0.45\linewidth]{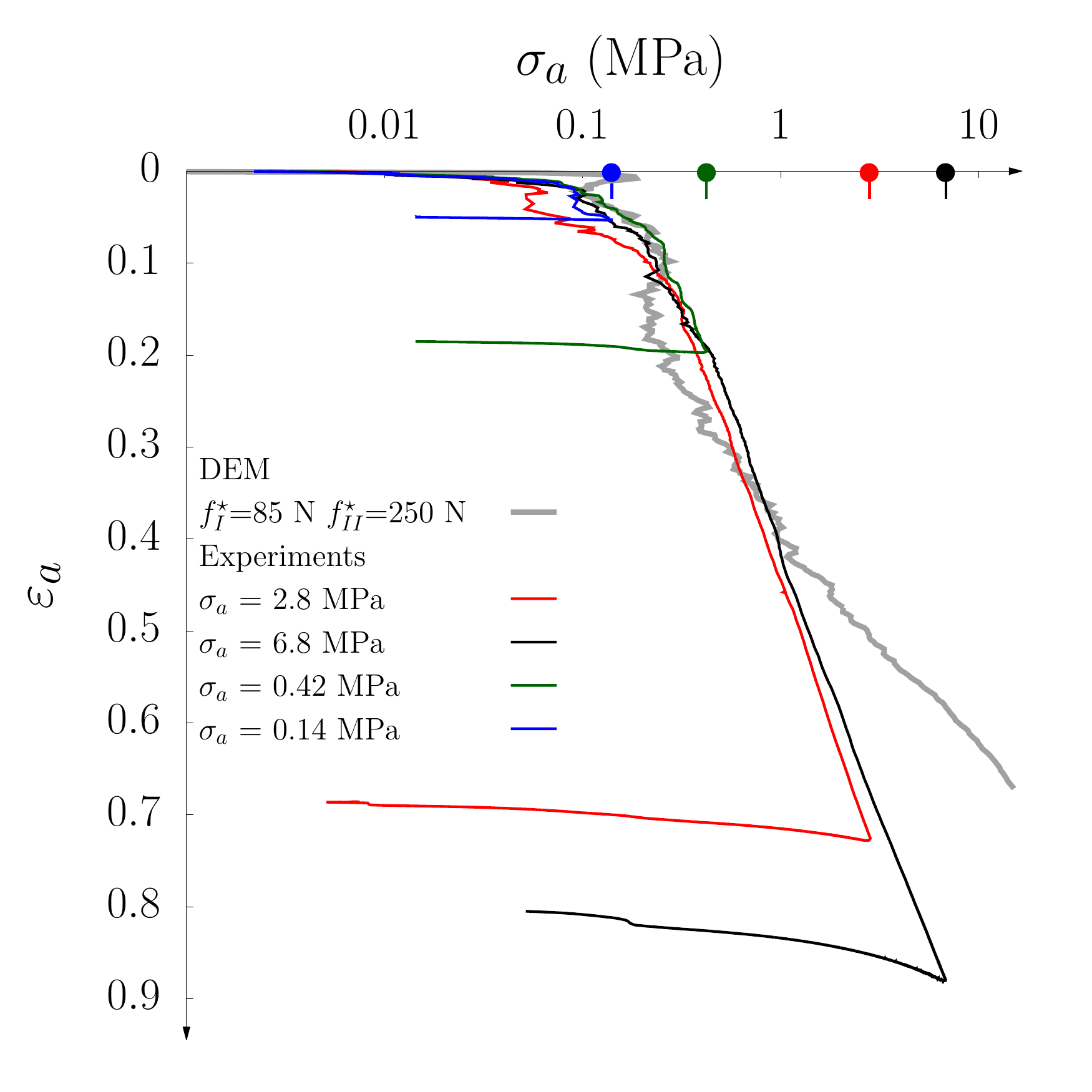}
\begin{flushleft} \hspace{4cm} \textbf{(a)} \hspace{8cm} \textbf{(b)} \end{flushleft}
\caption{Uniaxial compression of brittle hollow particles. (a) grain size distribution (GSD) of a DEM specimen of clusters compared with a real sample made of clay-shells at different stages of oedometric compression. Dots joined by continuous lines present the results of the sieve analysis \cite{ReportCermesOedo} for experimental tests stopped at various stress stages -- theses stages are identified on the response curves (b). For the simulations, the sieving being computerised, the hatched area gives an estimate of the GSDs.}
\label{FigGradingCompare}
\end{figure*}

In the previous sections, we presented the influence of various numerical and geometrical parameters on the numerical macroscopic behaviour of the clusters specimens. A broad experimental campaign has been performed by collaborators of \emph{Navier} Laboratory in Paris \cite{ReportCermesOedo}, to study the mechanical behaviour from samples made of around 2\,000 clay-shells {of identical size}.
In each uniaxial compression test, the evolution of stress and strain is characterized by particle rearrangements but also, and more importantly, by particle breaking -- both in experiments and in DEM simulations. Therefore, we propose to perform a DEM-experiment confrontation at the particle scale to compare the evolution of grain breakage {of initially identical cluster size} throughout the uniaxial compression test. In the Figure~\ref{FigGradingCompare}b, 5 strain-stress curves of oedometer tests are presented: one obtained from a DEM computation (gray curve); 4 experimental tests stopped at various stress levels (blue, $\sigma_a = 0.14$~MPa -- green, $\sigma_a = 0.42$~MPa -- red, $\sigma_a = 2.8$~MPa -- black, $\sigma_a = 6.8$~MPa). For each of these 4 stress levels, a grading analysis were performed for the 4 experiments as for the DEM simulation. Experimentally, French standard sieves were used (mesh sizes of 3.15~mm, 5~mm, 12.5~mm and 16~mm). The grading curves of the 4 samples corresponding to 4 stress levels are shown of the Figure~\ref{FigGradingCompare}a with circular points connected by lines. The grading curves of the broken clusters in the simulation differ from the experimental ones because sieves are not needed and the smallest passing size is perfectly known. Once broken, cluster pieces hereafter called \emph{fragments} can be made of 
1 to 11 segments, the intact clusters being made of 12 segments. Assuming that all fragments pass the sieve perfectly vertically, each fragment is associated to the dimension of its annular cross section.  Therefore, six types of fragment can be distinguished, and then sizes: 1 to 5 sectors and sectors with 6-11 sectors that have a sieve size identical to that of the intact cluster.
Unlike clay-shell fragments, the sizes of cluster fragments are perfectly known. We can therefore assume that a sieve with a mesh size exactly equal to that of the fragment would allow the fragment to pass through, while a sieve with a slightly smaller mesh size would retain the fragment. Thus, for a sieve size corresponding to a given fragment size, we chose to consider that these fragments could pass through \emph{and} be retained by that sieve. Therefore, the particle size curve obtained from the analysis of the fragments produced during the DEM simulation is not a curve but an area (hatched part in the Figure~\ref{FigGradingCompare}a) bounded in its upper part by what passes through the sieve and in its lower part by what is retained by the sieve. Whereas the clay-shell particles can be powdered, the smallest fragment, a sector, is the smallest grain size for the discrete element model. Thus, it can now be observed in Figure~\ref{FigGradingCompare}a that the experimental grain size distributions (GSDs) obtained at different levels of vertical stress are in a good agreement with DEM (the dashed areas of the same color). This result is crucial in a sense that it gives confidence in the relevance of the experimentally identified values of $f^\star_I$ and $f^\star_{II}$ used for the DEM model.

\subsection{Setting the model parameters}

In the previous sections, all the numerical, mechanical and geometrical parameters (internal structure parameters) that can influence the macroscopic behaviour have been presented and tested. It was thus shown that, although the brittle grain model proposed here has a great amount of parameters, few of them seem to have a significative influence on the kinetics of grain breakage during sample deformation subjected to oedometric loading; in particular, the number of segments used to define a cluster is not a sensitive parameter for non-extreme compression, provided that a minimal number is used. 
The only really sensitive parameters are, at the grain scale, the mode-\textit{I} fracture-triggering parameter $f^\star_I$ (recall that $f^\star_{II}$ does depend on $f^\star_I$), and the Coulomb friction angle between the grains -- all other numerical parameters showed a relatively small influence. 
The other relevant parameters are, at the sample scale, the initial density of the sample (number of particles per cubic meter) and the initial orientation of the shell-particles.

Therefore, using a cylindrical specimen of about 2\,000 clusters, we propose to make a comparison with a real sample of about 2\,000 clay-shells, both subjected to oedometric compression. For clusters as for the clay-shells, the grains are randomly oriented during the packing preparation. the set of numerical parameter used for the DEM oedometric compression is summarized in Table~\ref{TabFinalParam}.

\begin{table}[htb]
\centering
\begin{tabular}{ll | ll}
\toprule
\multicolumn{4}{l}{1.~Force laws}\\
\cmidrule(lr){1-4}
\multicolumn{2}{l|}{1.1~Links} &  \multicolumn{2}{l}{1.2~Frictional contacts}\\
\cmidrule(lr){1-4}
$k_{I}$ & 5.5~$10^{6} $ N/m & $k_{n}$ & 5.5~$10^{6}$  N/m \\
$k_{II}$ & 5.5~$10^{6}$  N/m& $k_{t}$ & 5.5~$10^{6}$  N/m \\
$f_{I}^\star$ & Weibull for $m=7.2$, $x_0=190$~N & $\mu_{shell}$ & 0.30 \\
$f_{II}^\star$ &  $\text{f}(f_{I}^\star)=f_{I}^\star/0.34$  & $\mu_{wall}$ & 0.15\\
$q$ & 2 & &  \\
\cmidrule(lr){1-4}
\multicolumn{4}{l}{2.~Shape discretisation}\\
\cmidrule(lr){1-4}
$N^\star_\text{circ}$ & 12 & $N^\star_\text{axial}$ & 1\\
\cmidrule(lr){1-4}
\multicolumn{4}{l}{3.~Initial density}\\
\cmidrule(lr){1-4}
$n$ & \multicolumn{3}{l}{140\: 388~m$^{-3}$ (loose)}  \\
\bottomrule
\end{tabular}
\caption{Summary of the parameters used for the oedometric compression (see Figure~\ref{FigFinalComCurve}).}
\label{TabFinalParam}
\end{table}

\begin{figure}[htb]
\includegraphics[width=0.99\linewidth,trim={0 0 20 0},clip]{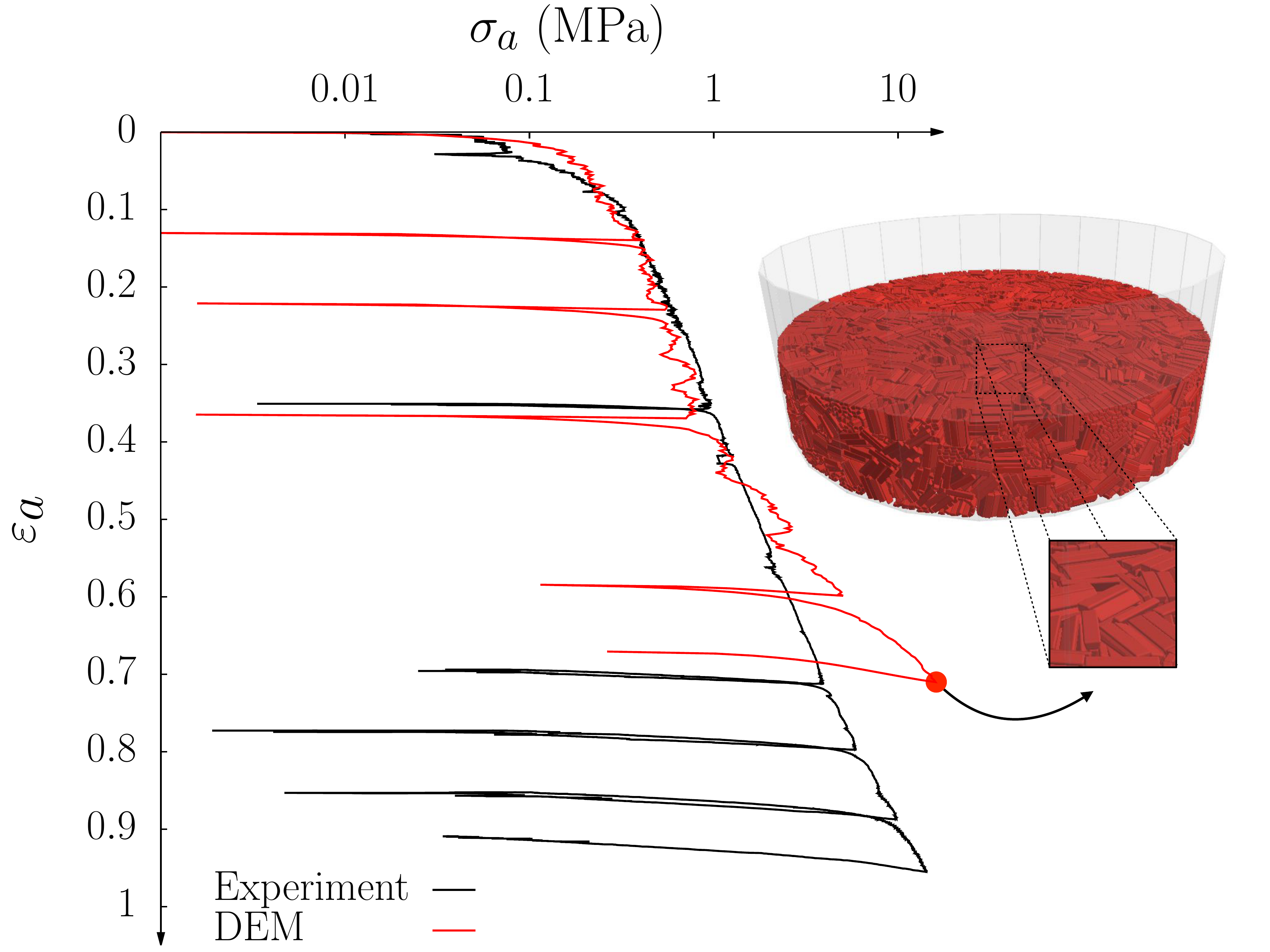}
\caption{Oedometer compression test. Comparison of macroscopic stress-strain curve. Sample size: $D=35$~cm and $H_0\approx13$~cm. DEM simulations on $\sim2\:000$ clusters for a set of parameters given in Table~\ref{TabFinalParam}. The experiment performed at \textit{Euro-G\'eomat-Consulting EGC} \cite{RaportOedoEGC}. The axial strain $\varepsilon_a$ is computed using Eq.~(\ref{EqStrain}).}
\label{FigFinalComCurve}
\end{figure}

Figure~\ref{FigFinalComCurve} compares the outcome of the DEM modelling with the experimental $\sigma_a:\varepsilon_a$ curve. The experimental oedometer test performed by \textit{Euro-G\'eomat-Consulting} EGC \cite{RaportOedoEGC} includes series of the unloading-reloading (UR) cycles -- UR cycles were also performed on the DEM model. 
With the experiments, it is noticeable that the clay-shells assembly subjected to an oedometer-type loading path can reach a compression level close to a vertical strain $\varepsilon_a  = 1$. At this stage of deformation, the clay-shells are totally crushed and reduced to powder. If the compression test had been continued, the stress would have increased sharply, as the sample would hardly increase its vertical strain. 
With a DEM simulation, the clusters cannot be reduced to powder, the smallest element of the model being the sector. Consequently, 
the sample stiffness will tend to the sector-to-sector stiffness for smaller axial strain.
Thus, while the two experimental and numerical $\sigma_a$\ding{214}$\varepsilon_a$ curves are initially very similar, they diverge for $\varepsilon_a \simeq 0.45$. At $\varepsilon_a \simeq 0.45$, the numerical curve presents an \emph{inflexion point}. An exponential law \cite{Bauer1996,Hu2011} seems well suited to describe the trend of the classic compression curve, \textit{i.e.}, for $\varepsilon_a \le 0.45$. In the case of tests with a large strain range, those laws loose their validity after the inflexion point. On the experimental curve the inflexion point is less clearly marked, but it appears at $\varepsilon_a \simeq 0.6$ such that oedometric modulus rises ($\sigma_a$ rises up faster).

Our discrete element model limits $\varepsilon_a$ to about 0.7. This limitation results from the large size of the sector chosen (sectors are as long as the tube shape), because the tube shape was partitioned only circumferentially: $N^\star_\text{circ}= 12$ and  $N^\star_\text{axial}= 1$.  Increasing $N^\star_\text{axial}$, as shown in the Figure~\ref{FigShellS2}, would not only have widened the range of the maximum $\varepsilon_a$ but also strain-postponed the inflexion point. On the contrary, we checked that increasing  $N^\star_\text{circ}$ brings no benefit as long as $N^\star_\text{axial}$ remains unchanged. 
By reducing the number of sectors with $N^\star_\text{circ}= 12$ and $N^\star_\text{axial}= 1$, we intended to limit the computational time (a DEM simulation of a compression test on 2\,000 clusters from $\varepsilon_a =0$ to $\varepsilon_a =0.7$ runs during 6 to 7 weeks on 6 cores).

The good agreement observed between the DEM and experiments is mainly due to the choice of the parameters $f_I^\star$ and $f_{II}^\star$, $\mu$ and the initial density of the sample. The contact elasticity parameters $k_{I}$ and $k_{II}$ play only a very limited role in the observed behavior, as long as the specimen is monotonically loaded. To justify the correct choice of elasticity parameters in the model, we have, as in the experiment, carried out UR cycles, which can be considered as quasi-elastic and therefore very dependent on the elasticity parameters of the contact between the particles.  However, as it can be seen in Figure~\ref{FigFinalComCurve}, the UR moduli obtained \textit{via} the DEM are equivalent to those observed experimentally -- the contact elasticity parameters have therefore been correctly chosen. This validates these parameters but also the procedure used to determine them (see Section~\ref{secPramMicro}). 

\section{An analytical model for compression-strain arising from shell comminution\label{secDiscus}}

\subsection{Grain breakage analysis}

In Section~\ref{secPramMicro}, we introduced a distinction between two breakage regimes: the cluster breakage is said to be \emph{primary} when an intact cluster breaks and consequently make its internal void accessible. The breakage is called \emph{secondary} when parts of already broken cluster can continue to break
without ``releasing'' too many voids.
In primary breakage, when voids become physically accessible, they play a key role in the \emph{compressibility} of the whole granular specimen. Note that we refer to compressibility in terms of void reduction, exclusively (cluster sectors being unbreakable and assumed perfectly rigid).

\begin{figure}[htb]
\centering
\includegraphics[width=0.98\linewidth,trim={0 50 20 20},clip]{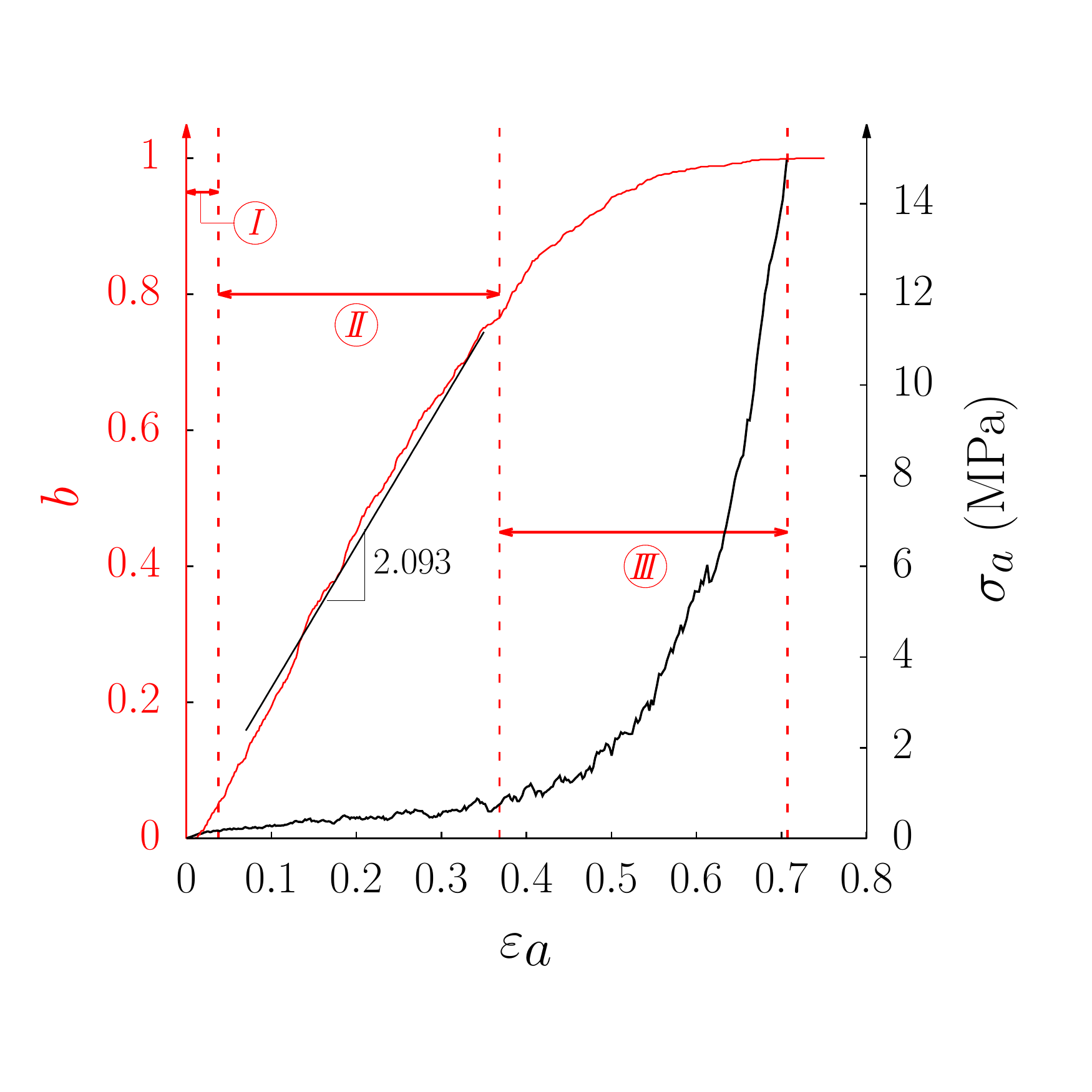}
\caption{Influence of the rate of broken clusters \textit{b} on the mechanical behaviour for the loose sample. Three phases can be distinguished for the axial strain as a reference: \textcircled{\scriptsize \textit{I}} the lack or the onset of breakage, \textcircled{\scriptsize \textit{I\!I}} a linear growth, and \textcircled{\scriptsize \textit{I\!I\!I}} a boosted increase of the vertical stress.}
\label{FigBreakageStrain}
\end{figure}

To express the level of primary breakage, a simple ratio can be used:
\begin{equation} 
b = \frac{N_\text{broken}}{N_\emptyset}
\label{EqDemage}
\end{equation}
where $N_\emptyset$ is the initial number of clusters in the specimen. The number of primarily broken clusters $N_\text{broken}$ increases from the beginning of the oedometric compression and cannot exceed $N_\emptyset$. Hence, $b$ ranges by definition between 0 and 1; it is a proportion of broken clusters.
In Figure~\ref{FigBreakageStrain}, the evolution of $b$ (red curve) compared to the stress-strain response (black curve) clearly demonstrates that cluster breakage influences the mechanical behaviour.

The evolution of $b$ with respect to $\varepsilon_a$ shows 3 phases: in the \textcircled{\scriptsize \textit{I}} phase, the sample compresses mainly due to cluster rearrangements with little cluster breakage ($b<0.05$). At some point, the clusters can no longer rearrange, higher inter-cluster forces develop, and the particles are more prone to break up. This corresponds to the beginning of the second phase, \textcircled{\scriptsize \textit{I\!I}}.
Then, the breakage occurs rapidly with an approximately constant slope of {2.093 with the increase of $\varepsilon_a$}. At the same time, the macroscopic stress increases slightly and linearly as a function of the axial strain.
The phase \textcircled{\scriptsize \textit{I\!I}} begins when the mechanical behavior curve becomes nonlinear, even exponential-like, for $b \ge 80\%$.  In this phase, the higher the breakage rate $b$, the faster $\sigma_a$ increases. When all the clusters are broken ($b=1$), the internal voids of each cluster are released.

\subsection{An amended concept of the void ratio}

The void ratio $e$ is commonly used to describe the volume changes of compressible soils. It is a dimensionless parameter measuring the voids as a fraction of the solid phase:
\begin{equation} 
e=V_\mathrm{v}/V_\mathrm{s}=(V_\mathrm{tot}-V_\mathrm{s})/V_\mathrm{s}
\label{EqVoid}
\end{equation}
where the solid volume is a sum of cluster volume $V_\mathrm{s}^i$ (being itself the sum of the volumes of sectors composing the cluster) over all the clusters $N_\emptyset$:
\begin{equation} 
V_\mathrm{s}=\sum_{i=1}^{N_\emptyset} \: V_\mathrm{s}^i
\label{EqVolumeSolid}
\end{equation}

\begin{figure}[htb]
\begin{flushleft} \hspace{1.35cm} (a) \hspace{2.25cm} (b) \hspace{2.25cm} (c)  \end{flushleft}
\centering
\includegraphics[width=0.32\linewidth]{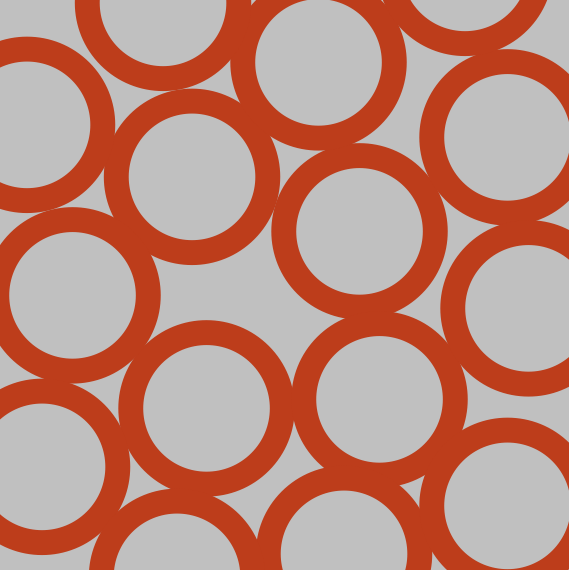}
\includegraphics[width=0.32\linewidth]{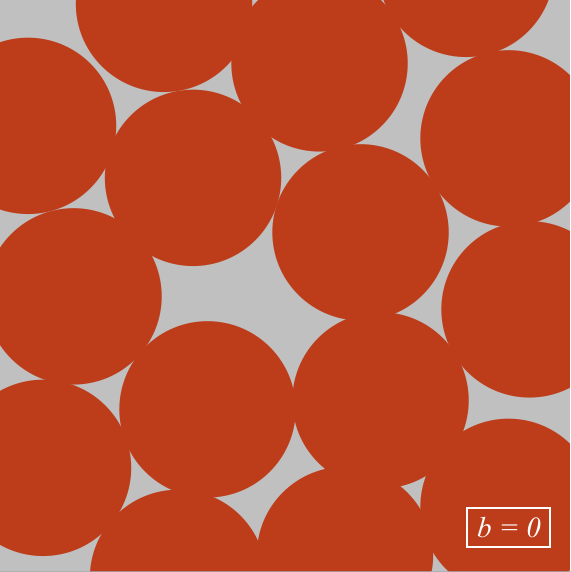}
\includegraphics[width=0.32\linewidth]{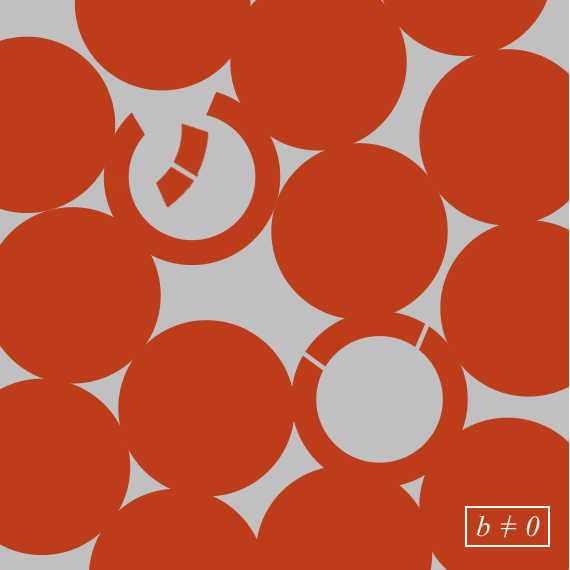}
\caption{An 2D illustrations of partition into voids (grey) and solid (red) phases. (a) Most common definition of the void ratio $e$ where the void phase is the overall air and the solid phase is the clay matter. The modified void ratio $e^\star$ (the inter-granular void ratio) accounts for the accessibility to the internal voids: (b) all shells are intact ($b=0$) and (c) some shells are broken ($b>0$) making their inside a part of the void phase. }
\label{FigVoidRatioCalc}
\end{figure}

Figure~\ref{FigVoidRatioCalc}a shows the classical division of total volume $V_\text{tot}$ into two sub-volumes: the voids $V_\text{v}$ (grey) and the solid phase $V_\text{s}$ (red). Within an assembly of intact clusters, the hollow space inside the tubes is inaccessible by other clusters. This geometric exclusion implies that the space $V_\text{v}^i$ trapped inside the intact clusters $i$ can be assigned to the solid phase as presented in Figure~\ref{FigVoidRatioCalc}b. When the cluster breaks, its internal void area becomes accessible and is considered as a free space (Figure~\ref{FigVoidRatioCalc}c). 
Hence, we suggest to amend the definition of $e$ into a suitable void ratio $e^\star$ that accounts for the evolution of accessible voids into inaccessible voids:
\begin{equation} 
e^\star
=\frac{V_\text{accessible}}{V_\text{inaccessible}}
=\frac{V_\text{tot}-(V_\text{s}+V^\star)}{V_\text{s}+V^\star}
\label{EqVoidMix}
\end{equation}
where $V_\text{accessible}$ and $V_\text{inaccessible}$ are the whole volumes that can or cannot be filled by matter, respectively. To calculate $e^\star$ the breakage ratio $b$ is introduced in the calculation of the internal inaccessible voids treated as the solid:
\begin{equation}\label{EqVstar}
V^\star = (1-b) \sum_{i=1}^{N_\emptyset}V_\mathrm{v}^i
\end{equation}

\begin{figure}[htb]
\centering
\includegraphics[width=0.98\linewidth]{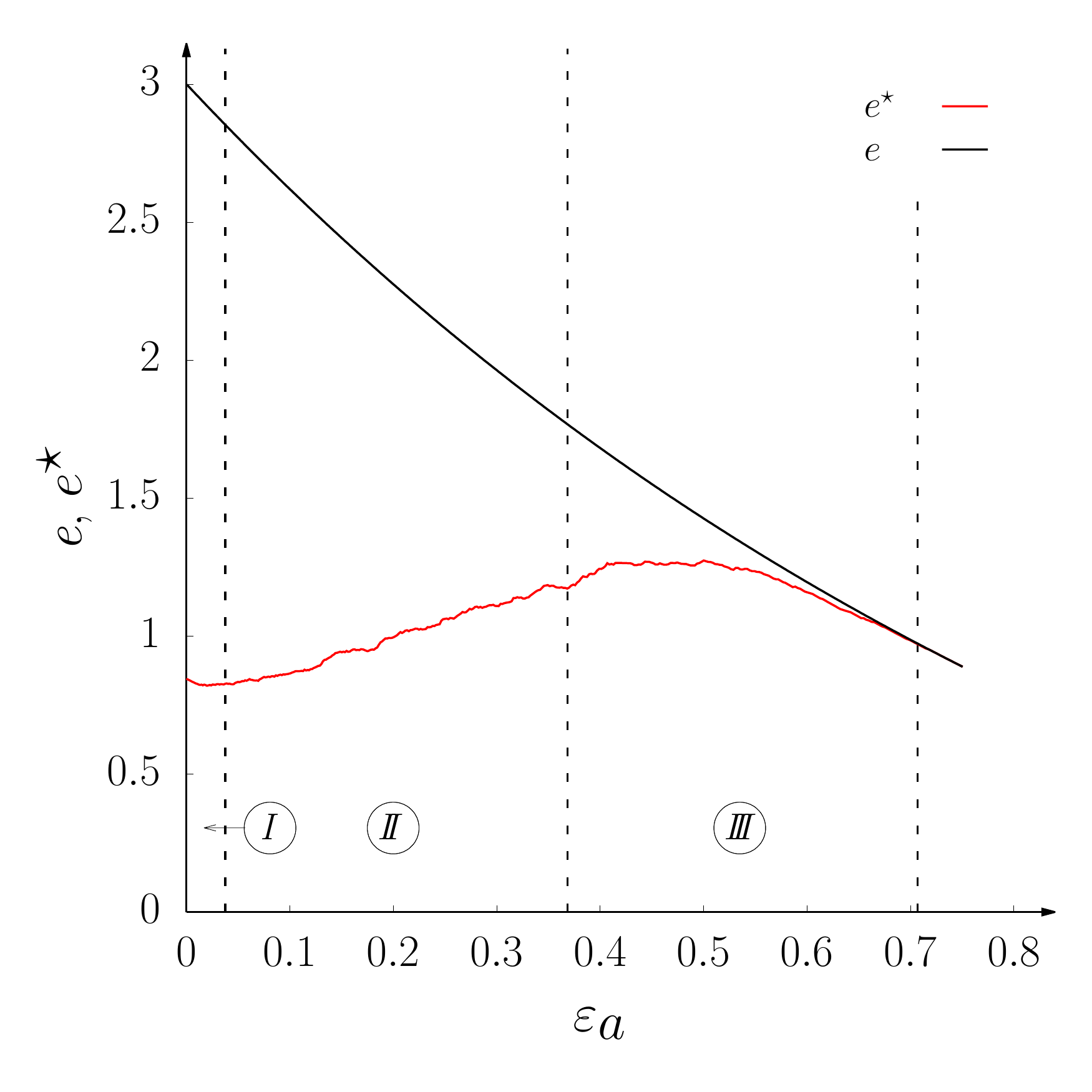}
\caption{The compression curves with respect to axial strain. Evolution of the standard void ratio $e$ (Eq.~\eqref{EqVoid}) and of the modified void ratio $e^\star$ (Eq.~\eqref{EqVoidMix}) in the course of uniaxial compression at constant velocity rate.
This simulation is the same as the one shown in Figure~\ref{FigFinalComCurve}, with the parameters of Table~\ref{TabFinalParam}.}
\label{FigVoidRatioStrain}
\end{figure}

Figure~\ref{FigVoidRatioStrain} shows the evolution of both $e$ and $e^\star$ as a function of the axial strain $\varepsilon_a$, computed with Eqs.~\eqref{EqVoid} and \eqref{EqVoidMix}.
As already mentioned, $\varepsilon_a$ was defined in the framework of the natural strain, that is calculated using a logarithmic function (Eq. \eqref{EqStrain}). Therefore, $e$ always decreases non-linearly as a function of $\varepsilon_a$. 
On the contrary, $e^\star$ evolves in a non-monotonous manner. In the phase \textcircled{\scriptsize \textit{I}},
$e^\star$ remains around its initial value $e^\star_0$ because of the clusters rearrangements and the onset of breakage. 
A rather stable increase of $e^\star$ is observed in the phase \textcircled{\scriptsize \textit{I\!I}} of extensive breakage. 
This is a consequence of many voids becoming prone to the progressing overall irreversible compression (decrease of $V^\star$).
The curve exposes the breakage inhibition with a maximum value of $e^\star$ (for $\varepsilon_a \simeq 0.42$ in Figure~\ref{FigVoidRatioStrain}).
It corresponds to a maximum of $e^\star$ after which the sample becomes denser. This maximum indicates the axial strain at which the highly compressible regime ends. Once all the clusters are crushed, ($b=1$, or equivalently $V^\star=0~\text{m}^3$) the two curves must merge because  Eqs.~\eqref{EqVoidMix} and \eqref{EqVoid} become identical.

The control of the maximum of $e^\star$ enables to tune the compressible capacities of the cluster assembly. This opens the door to the optimisation of the material strength and the cluster geometry. 
To this end, a prediction of the $e^\star$\ding{214}$\varepsilon_a$ relationship is proposed.
By dividing all terms of Eq.~\eqref{EqVoidMix} by $V_\mathrm{s}$, the modified void ratio can be expressed as a function of the standard void ratio $e$:
\begin{equation} 
e^\star=\frac{e-(V^\star/V_\mathrm{s})}{1+(V^\star/V_\mathrm{s})}
\label{EqVoidMixStep2}
\end{equation}
Using Eqs.~\eqref{EqVolumeSolid} and \eqref{EqVstar}, the ratio $V^\star/V_\mathrm{s}$ becomes  
\begin{equation} 
\frac{V^\star}{V_\mathrm{s}} = \left(1-\dfrac{N_\mathrm{broken}}{N}\right) \frac{V_\mathrm{v}^i}{V_\mathrm{s}^i} = (1-b)E_0
\label{EqComponent}
\end{equation}
where the constant $E_0={V_\mathrm{v}^i}/{V_\mathrm{s}^i}$ is the cluster scale void ratio. 
As a consequence, combining Eqs.~\eqref{EqVoidMixStep2} and \eqref{EqComponent} lead to
\begin{equation} 
e^\star(e,b) = 
\frac{e-(1-b)E_0}{1+(1-b)E_0}
\label{EqPredVoid}
\end{equation}
where $e^{\star}$ depends on the void ratio $e$, the initial void ratio $E_0$ of the intact clusters, and the evolution of the rate of breakage $b$ during the compression.

To express $e^{\star}$ as a function of $\varepsilon_a$, both $b$ and $e$ need to be a related to $\varepsilon_a$. Firstly, $b$  \ding{214} $\varepsilon_a$ relationship was assumed to be linear with slope of {$\simeq 2$ (which corresponds to the slope $2.093$ shown in Figure~\ref{FigBreakageStrain}}):
\begin{equation}
{b = 2 \: \varepsilon_{a}}
\label{Eqbversusstrain}
\end{equation}
{with the logarithmic strain definition in uniaxial compression defined as}
\begin{equation}
\varepsilon_{a}=\ln\left(\frac{V_{\text{tot},\emptyset}}{V_\text{tot}}\right) 
\label{EqLargeStrainStep2}
\end{equation}
where subscript $\emptyset$ denotes the initial state. We can then write that
\begin{equation}
\mathrm{exp}(\varepsilon_a)=\frac{e_0+1}{e+1}
\label{EqLargeStrainStep3}
\end{equation}
and finally, 
\begin{equation}
e(\varepsilon_a) = \frac{1+e_0}{\exp(\varepsilon_a)} - 1
\label{EqLargeStrainStep4}
\end{equation}
with the initial void ratio $e_0$ serving as the input.

\begin{figure}[htb]
\centering
\includegraphics[width=0.9\linewidth]{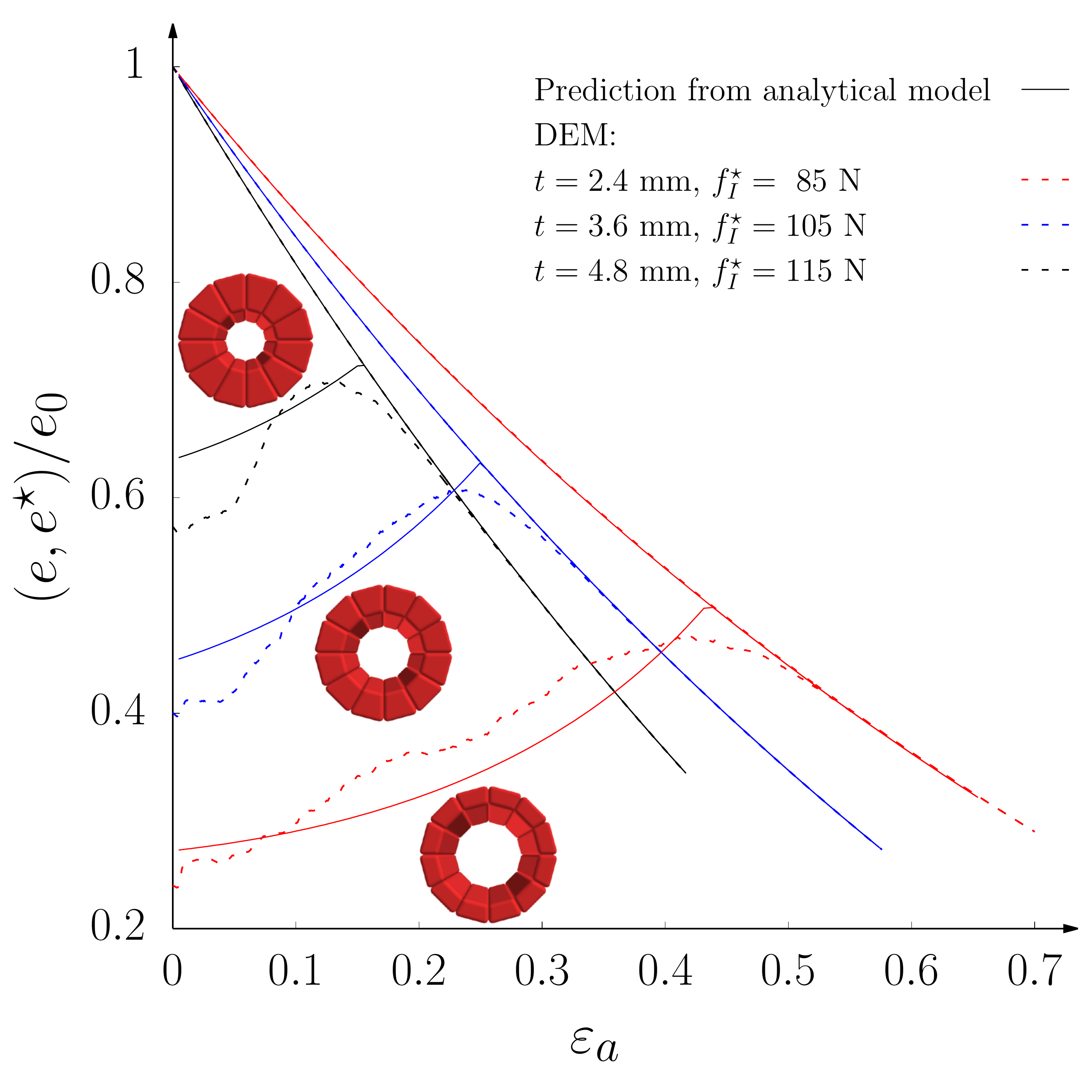} 
\caption{Numerical evolution of normalised void ratios $e/e_0$ and $e^{\star}/e_0$ (dashed lines) with respect to axial strain. Predicted compression curves (solid lines) according to equation~\eqref{EqPredVoid} for shells of various thickness $t$ with (red) $b=2 \: \varepsilon_a$ for $t = 2.4$~mm,  (blue) $b=4 \: \varepsilon_a$ for $t = 3.6$~mm, and (black) $b=6.5 \: \varepsilon_a$ for $t = 4.8$~mm. These prediction are compared with the DEM modelings (dashed lines).}
\label{FigPredStrainThick}

\end{figure}

In summary, the model distinguishes two regimes depending on the breakage level: 
Eqs.~\eqref{EqPredVoid}, \eqref{Eqbversusstrain} and \eqref{EqLargeStrainStep4} constitute the prediction of $e^{\star}$ for $b < 1$. Once all the clusters are broken, $b=1$, Eq.~\eqref{Eqbversusstrain} is no longer valid and the classical definition of the void ratio is used. Hence, the model can be written as:
\begin{equation}
\begin{cases}
\text{for}~0\leq b<1 & \displaystyle e^\star(e,b) = \frac{e-(1-b)E_0}{1+(1-b)E_0}\\
\text{for}~b=1 & e^\star=e
\end{cases}
\label{EqSumModel}
\end{equation}

\subsection{Testing the model}

Figure~\ref{FigPredStrainThick} shows the outcome of the prediction model for different thicknesses of shells ($t = 2.4;\ 3.6;\ 4.8$~mm, which corresponds to $E_0 \simeq 1.062;\ 0.498;\ 0.234$, respectively). These predictions (solid lines) are also compared with the equivalent simulations (dashed lines).

The model was built upon a very basic concept and a simple assumption on breakage kinematics, Eq.~\eqref{Eqbversusstrain}.
Although the model is unable to replay all the details in the evolution of the $e^\star$, it captures quite well the main features, especially when the void ratio switches from $e^{\star}$ to $e$.
In essence, this model has structured a solid foundation for a more complex prediction of the constitutive  $e^\star$\ding{214}$\sigma_a$ relationship to be derived in the future.

\section{Summary\label{secEnd}}

This paper gives some insight into the micro-mechanical origins of the high compressibility of the assembly that comprises extremely porous grains. Hereinbefore, we broke down the complexity of the model into several steps and build it up from micro to macro scale, as follows.

In the Section~\ref{secModel}, by bonding non-spherical particles (called sectors), we have designed a discrete element (DE) model of a tubular grain (\textit{i.e.}, a cylindrical shell) capable of simulating brittle crushing.
An improvement of the rigid clustering technique that allows efficient modeling of particles of any complex shape has been introduced. To our knowledge, clustering \emph{sphero-polyhedral} elements into breakable grain is a rather innovative idea in the field of crushable geo-materials. The necessary numerical implementation were included in the in-house code \texttt{Rockable}.

The tensile failure has been achieved at the shell-scale using uniaxial radial compression (URC) in the Section~\ref{secPramMicro}. The variability of the shell strength was described through a Weibullian distribution. The experimental campaign contributed to numerical study not only with the determination of the tensile strength and its variability, but also with a characterisation of the breakage manner, helpful to construct the geometry of the cluster. 
The model successfully captured the shell mechanical behaviour in URC under the assumption of the linear elasticity. 
Still, the first consequences of the model simplifications have been identified due to the lack of (\textit{i}) non-linearity of the contacts laws (delaying the arise of intra-cluster forces) and (\textit{ii}) separation between the crack initiation and propagation. A parametric study has been undertaken to clarify the influence the elastic and yielding parameters. The number of sectors per cluster have been reported to affect the values of discrete parameters. Finally, a robust set of parameters and a cluster structure tradeoff were selected.

Section~\ref{secExpeCharacterisation} comments on the ``structure'' of actual specimens. Acknowledging the influence of initial state on the mechanical behaviour, we characterised the bulk density and the intrinsic anisotropy of shell orientations. While the former was a simple and straightforward measure, the latter required the use of more advanced techniques.
First, we have proposed a geometry-based approach to efficiently detect the tubes in 3D binary image resulting from X-ray scanning. It was then followed by the statistical analysis of the orientations. 
Despite a natural tendency of the shell to embed horizontally, the remaining orientations are uniformly distributed within the sample. As a consequence, under oedometeric compression, the tensile failure of shell will be dominant.

In the Section~\ref{secParamStudy}, a systematic investigation of the responsiveness of the macroscopic model to changes in the {DEM} parameters has been conducted. Beyond that, the sensitivity analysis has been extended to the influence of the initial state and of the cluster geometry. To moderate the mechanical response numerically, two parameters have been found crucial.
First, it has been shown that the axial stress, that is the sample bearing capacity, depends on the tensile strength of constituents which is controlled by the contact parameter called the tensile yield force (see Section~\ref{secModel}). Second, the initial density shapes (either smoothen or sharpen) the trend of the constitutive relationship. 
It has been clearly demonstrated that the significant size of the sector in longitude size of longitude sector restricted the range of the strain. On the other hand, the evolution of the grading size distribution highlighted that the micro-mechanics of packing has not been harmed with this simplification. Decrease of the sector size extends the functionality of the model by postponing the end-deformation of the primary breaks, but becomes computationally expensive.
The effort undertaken has made it possible to achieve an important objective: to reproduce the experimental result provided by EGC and Laboratoire Navier, permitting to satisfy the technological specification from fabricating the shell, to the prefabrication of segment lining.

The Section~\ref{secDiscus} evidenced that the high compressibility originated from the extensive internal pore collapse triggered by the primary breakage. The standard void ratio has been redefined in the framework of accessible and inaccessible space, instead of solid and void. 
The overall void ratio decreased progressively respecting the stain-controlled compression. In contrast, the redefinition of porosity taking into account the accessibility of voids has led to a strong change in this evolution. 
The accessible void ratio increases despite the fact that the macroscopic volume shrink. 
In other words, the packing experienced a ``porosification'', and once the extensive breakage has been finished, only densification took place (\textit{i.e.}, a limited rearrangement of rigid sectors allowing high transmission of forces through the elastic stiff contacts). The end of this porosification was associated with the point at which the high compressibility could not be exploit any longer.
A general, geometrical development of a prediction model, including some analytically determined relationships, has been conducted to foresee how far in the macroscopic compression the internal porosification take place. In the comparison with modelling of the thicker shells, the estimations were rough but robust. This novel concept for compressibility analysis is then left as the basis for next upgrades and developments of the authors.

\section*{Acknowledgments\label{acknow}}

The Laboratoire 3SR is part of the LabEx Tec 21 (Investissements d'Avenir, Grant Agreement No. ANR-11-LABX-0030) 

\section*{Fundings and Conflicts of Interest}

This study was funded by ANDRA, the French National Radioactive Waste Management Agency. M. Stasiak, G. Combe and V. Richefeu received grants from ANDRA.  G. Armand and J. Zghondi are ANDRA's employees.



\end{document}